\documentclass[review,number,sort&compress]{elsarticle}

\usepackage{enumerate,booktabs,amssymb}
\usepackage{threeparttable}
\usepackage{bm,mathrsfs,bbm,amscd}
\usepackage[fleqn]{amsmath}
\usepackage{graphicx}
\usepackage{subfigure}
\usepackage{multirow}
\usepackage{multicol}
\usepackage[colorlinks=true]{hyperref}
\hypersetup{colorlinks = true,linkcolor = blue,anchorcolor =red,citecolor = blue,filecolor = red,urlcolor = red,pdfauthor=author}
\usepackage{upgreek,fancyhdr}
\journal{XX-XX}

\begin{document}

\begin{frontmatter}

\title{Construction and On-site Performance of the LHAASO WFCTA Camera}


\author[my26add,my27add]{F. Aharonian}
\author[my4add,my5add]{Q. An}
\author[my20add]{Axikegu}
\author[my21add]{L.X. Bai}
\author[my1add,my3add]{Y.X. Bai}
\author[my15add]{Y.W. Bao}
\author[my10add]{D. Bastieri}
\author[my1add,my2add,my3add]{X.J. Bi}
\author[my1add,my3add]{Y.J. Bi}
\author[my23add]{H. Cai}
\author[my10add]{J.T. Cai}
\author[my1add,my2add,my3add]{Z. Cao}
\author[my4add,my5add]{Z. Cao}
\author[my16add]{J. Chang}
\author[my1add,my3add,my4add]{J.F. Chang}
\author[my1add,my3add]{X.C. Chang}
\author[my13add]{B.M. Chen}
\author[my21add]{J. Chen}
\author[my1add,my2add,my3add]{L. Chen}
\author[my18add]{L. Chen}
\author[my20add]{L. Chen}
\author[my1add,my3add]{M.J. Chen}
\author[my1add,my3add,my4add]{M.L. Chen}
\author[my20add]{Q.H. Chen}
\author[my1add,my2add,my3add]{S.H. Chen}
\author[my1add,my3add]{S.Z. Chen}
\author[my22add]{T.L. Chen}
\author[my1add,my2add,my3add]{X.L. Chen}
\author[my15add]{Y. Chen}
\author[my1add,my3add]{N. Cheng}
\author[my1add,my3add]{Y.D. Cheng}
\author[my13add]{S.W. Cui}
\author[my7add]{X.H. Cui}
\author[my11add]{Y.D. Cui}
\author[my24add]{B.Z. Dai}
\author[my1add,my3add,my4add]{H.L. Dai}
\author[my15add]{Z.G. Dai}
\author[my22add]{Danzengluobu}
\author[my31add]{D. della Volpe}
\author[my28add]{B. D'Ettorre Piazzoli}
\author[my1add,my3add]{X.J. Dong}
\author[my10add]{J.H. Fan}
\author[my16add]{Y.Z. Fan}
\author[my1add,my3add]{Z.X. Fan}
\author[my24add]{J. Fang}
\author[my1add,my3add]{K. Fang}
\author[my17add]{C.F. Feng}
\author[my16add]{L. Feng}
\author[my1add,my3add]{S.H. Feng}
\author[my16add]{Y.L. Feng}
\author[my1add,my3add]{B. Gao}
\author[my17add]{C.D. Gao}
\author[my22add]{Q. Gao}
\author[my17add]{W. Gao}
\author[my24add]{M.M. Ge\corref{mycorrespondingauthor}}
\cortext[mycorrespondingauthor]{Corresponding author}
\ead{gemaomao@ynu.edu.cn}
\author[my1add,my3add]{L.S. Geng}
\author[my6add]{G.H. Gong}
\author[my1add,my3add]{Q.B. Gou}
\author[my1add,my3add,my4add]{M.H. Gu}
\author[my1add,my2add,my3add]{J.G. Guo}
\author[my20add]{X.L. Guo}
\author[my1add,my3add]{Y.Q. Guo}
\author[my1add,my2add,my3add,my16add]{Y.Y. Guo}
\author[my14add]{Y.A. Han}
\author[my1add,my2add,my3add]{H.H. He}
% ...

\author{(The LHAASO Collaboration)}

\address[my1add]{Key Laboratory of Particle Astrophyics \& Experimental Physics Division \& Computing Center, Institute of High Energy Physics, Chinese Academy of Sciences, 100049 Beijing, China}
\address[my2add]{University of Chinese Academy of Sciences, 100049 Beijing, China}
\address[my3add]{TIANFU Cosmic Ray Research Center, Chengdu, Sichuan,  China}
\address[my4add]{State Key Laboratory of Particle Detection and Electronics, China}
\address[my5add]{University of Science and Technology of China, 230026 Hefei, Anhui, China}
\address[my6add]{Department of Engineering Physics, Tsinghua University, 100084 Beijing, China}
\address[my7add]{National Astronomical Observatories, Chinese Academy of Sciences, 100101 Beijing, China}
% ...
\address[my10add]{Center for Astrophysics, Guangzhou University, 510006 Guangzhou, Guangdong, China}
\address[my11add]{School of Physics and Astronomy \& School of Physics (Guangzhou), Sun Yat-sen University, 519082 Zhuhai, Guangdong, China}
% ...
\address[my13add]{Hebei Normal University, 050024 Shijiazhuang, Hebei, China}
\address[my14add]{School of Physics and Microelectronics, Zhengzhou University, 450001 Zhengzhou, Henan, China}
\address[my15add]{School of Astronomy and Space Science, Nanjing University, 210023 Nanjing, Jiangsu, China}
\address[my16add]{Key Laboratory of Dark Matter and Space Astronomy, Purple Mountain Observatory, Chinese Academy of Sciences, 210023 Nanjing, Jiangsu, China}
\address[my17add]{Institute of Frontier and Interdisciplinary Science, Shandong University, 266237 Qingdao, Shandong, China}
\address[my18add]{Key Laboratory for Research in Galaxies and Cosmology, Shanghai Astronomical Observatory, Chinese Academy of Sciences, 200030 Shanghai, China}
% ...
\address[my20add]{School of Physical Science and Technology \&  School of Information Science and Technology, Southwest Jiaotong University, 610031 Chengdu, Sichuan, China}
\address[my21add]{College of Physics, Sichuan University, 610065 Chengdu, Sichuan, China}
\address[my22add]{Key Laboratory of Cosmic Rays (Tibet University), Ministry of Education, 850000 Lhasa, Tibet, China}
\address[my23add]{School of Physics and Technology, Wuhan University, 430072 Wuhan, Hubei, China}
\address[my24add]{School of Physics and Astronomy, Yunnan University, 650091 Kunming, Yunnan, China}
% ...
\address[my26add]{Dublin Institute for Advanced Studies, 31 Fitzwilliam Place, 2 Dublin, Ireland }
\address[my27add]{Max-Planck-Institut for Nuclear Physics, P.O. Box 103980, 69029  Heidelberg, Germany }
\address[my28add]{Dipartimento di Fisica dell'Universit\`a di Napoli   ``Federico II'', Complesso Universitario di Monte Sant'Angelo, via Cinthia, 80126 Napoli, Italy. }
% ...
\address[my31add]{D\'epartement de Physique Nucl\'eaire et Corpusculaire, Facult\'e de Sciences, Universit\'e de Gen\`eve, 24 Quai Ernest Ansermet, 1211 Geneva, Switzerland}
% ...

\begin{abstract}
	The focal plane camera is the core component of the Wide Field-of-view Cherenkov/fluorescence Telescope Array (WFCTA) of the Large High-Altitude Air Shower Observatory (LHAASO). Because of the capability of working under moonlight without aging, silicon photomultipliers (SiPM) have been proven to be not only an alternative but also an improvement to conventional photomultiplier tubes (PMT) in this application. Eighteen SiPM-based cameras with square light funnels have been built for WFCTA. The telescopes have collected more than 100 million cosmic ray events and preliminary results indicate that these cameras are capable of working under moonlight. The characteristics of the light funnels and SiPMs pose challenges (e.g. dynamic range, dark count rate, assembly techniques). In this paper, we present the design features, manufacturing techniques and performances of these cameras. Finally, the test facilities, the test methods and results of SiPMs in the cameras are reported here.
\end{abstract}


\end{frontmatter}

\section{Introduction} \label{sec:sec0}
The energy spectrum of cosmic rays (CRs) exhibits several interesting features that provide a wealth of information about their origin, acceleration, and propagation processes. One of the major scientific goals of the Large High-Altitude Air Shower Observatory (LHAASO)~\cite{caozhen1,caozhen3} is to explore these physical processes~\cite{kulikov1,blumera1} by measuring the energy spectrum and the composition of CRs. LHAASO consists of three ground-based detector arrays~\cite{hehuihai3}: an array of scintillator detectors and muon detectors covering an area of about 1~km$^{2}$ (KM2A), a Water Cherenkov Detector Array (WCDA), and a Wide Field-of-view Cherenkov Telescope Array (WFCTA) consisting of eighteen telescopes.

Energetic primary CR particles that enter the atmosphere induce extensive air showers (EAS)~\cite{grieder1}. EAS produce a large number of secondary photons which can be detected by the camera of the imaging atmosphere Cherenkov telescopes (IACT). The recorded image is then used to reconstruct the physical properties of the primary CR particles such as particle species, energy and incidence direction. The potential and efficiency of this method have been successfully demonstrated by previous experiments~\cite{baltrusaitis1,abbasi3,abraham1} and the WFCTA prototype experiment~\cite{zhangshoushan1,bartoli3}. WFCTA aims at measuring the spectrum of CRs from 10~TeV up to 1~EeV and studying its composition~\cite{zhangshoushan2}. WFCTA cannot be an effective gamma-ray detector compared to KM2A, but WFCTA, in conjunction with KM2A, provides cross calibration of the energy measurement of primary gamma-rays. WFCTA and KM2A adopt two independent methods to measure the energy of ultra-high energy gamma-rays in the energy range from 100 TeV to a few PeV, just as the AUGER experiment~\cite{auger1} provides two independent methods to measure ultra-high energy cosmic rays, namely the surface detector array~\cite{allekotte2008} and the fluorescence detector~\cite{abraham1}.

The main information about the telescopes of WFCTA can be found in Sec.~\ref{sec:sec1}. In Sec.~\ref{sec:sec2}, the WFCTA SiPM camera structure and main features are described. The laboratory test facilities, test methods, results, and assembly techniques of the camera are reported in Sec.~\ref{sec:sec3}. Finally, on-site running performance and on-site test results of these cameras are discussed in Sec.~\ref{sec:sec4}.

\section{The Telescope}\label{sec:sec1}
Each telescope of WFCTA consists of a segmented spherical mirror of about 5 m$^{2}$ with a SiPM-based camera installed at its focal plane. The light induced by EAS is focused on the camera by the mirror. The focal plane camera which has a field of view (FoV) of $16^{\circ}\times16^{\circ}$ is composed by an array of 32$\times$32 pixels, each with an angular size of $0.5^{\circ}\times0.5^{\circ}$, featuring a SiPM coupled to a square light funnel. The telescopes are movable, and can be operated in Cherenkov or fluorescence mode during different observation phases by changing the locations and the directions of the telescopes.

\begin{figure}[hbtp]
	\centering
	\includegraphics[width=12.0cm]{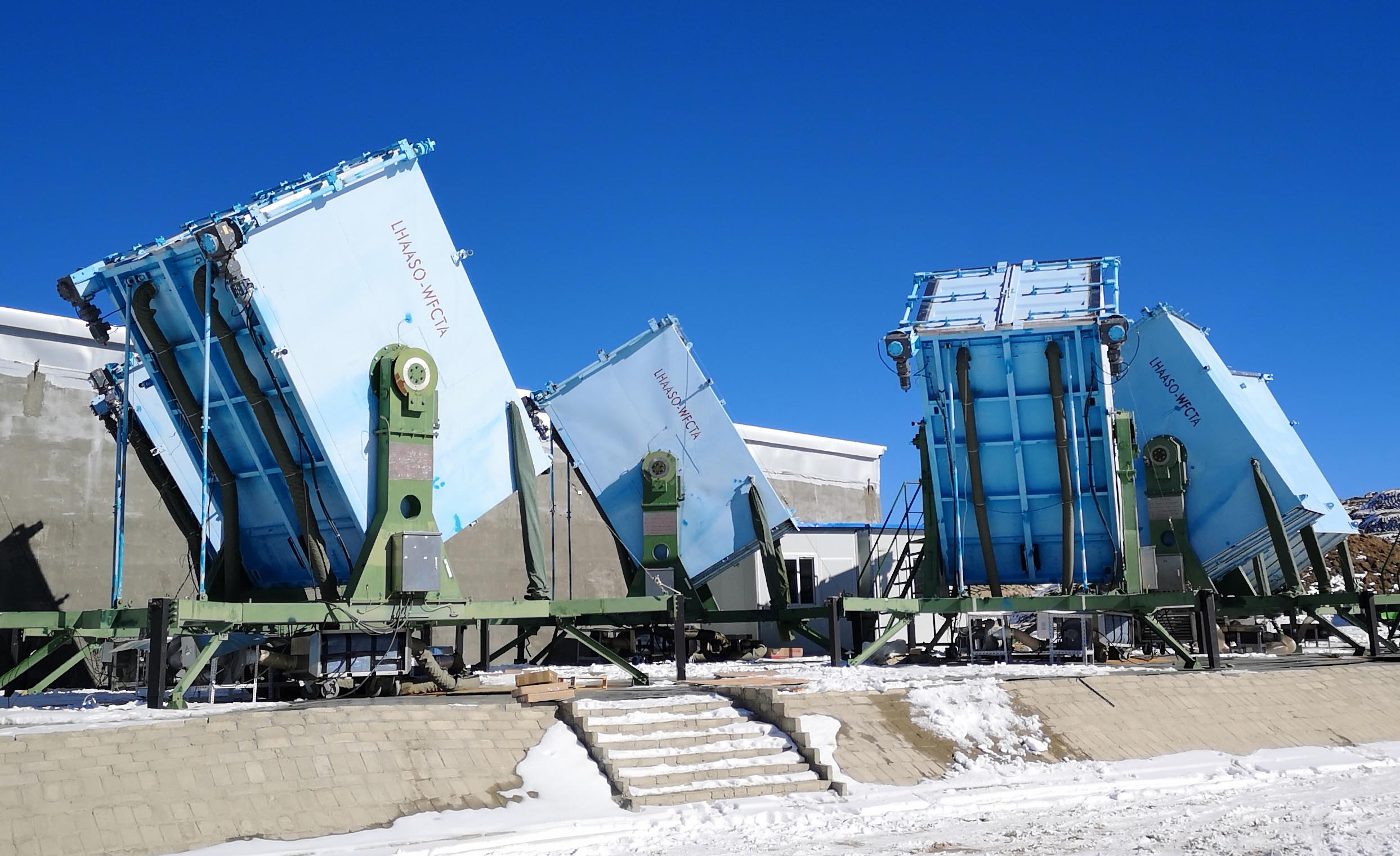}
	\caption{The photo of six telescopes of WFCTA.}
	\label{fig:telescops}
\end{figure}

Up to now, all of the telescopes have been deployed at the LHAASO site (E:$100^{\circ}03'$, N:$29^{\circ}18'$, 4410 m a.s.l.), which is located in Haizishan, Daocheng County, Sichuan Province, China. The photo of six telescopes is shown in Fig.~\ref{fig:telescops}.

\section{The SiPM Camera} \label{sec:sec2}
\subsection{Overview}
The SiPM camera converts optical signals generated by EAS into electrical signals, and records or discards them according to triggering criteria. The photon flux received by the camera is determined by the energy of the primary particle and the shower core distance. In each observation phase, the target energy span is about two orders of magnitude. Each pixel of the camera is required to accommodate a dynamic range from about 10 photo-electrons (p.e.) to 32,000 p.e. according to the results of the shower and the telescope simulation.

\begin{figure}[!htp]
	\centering
	\subfigure[]{\includegraphics[width=6.5cm]{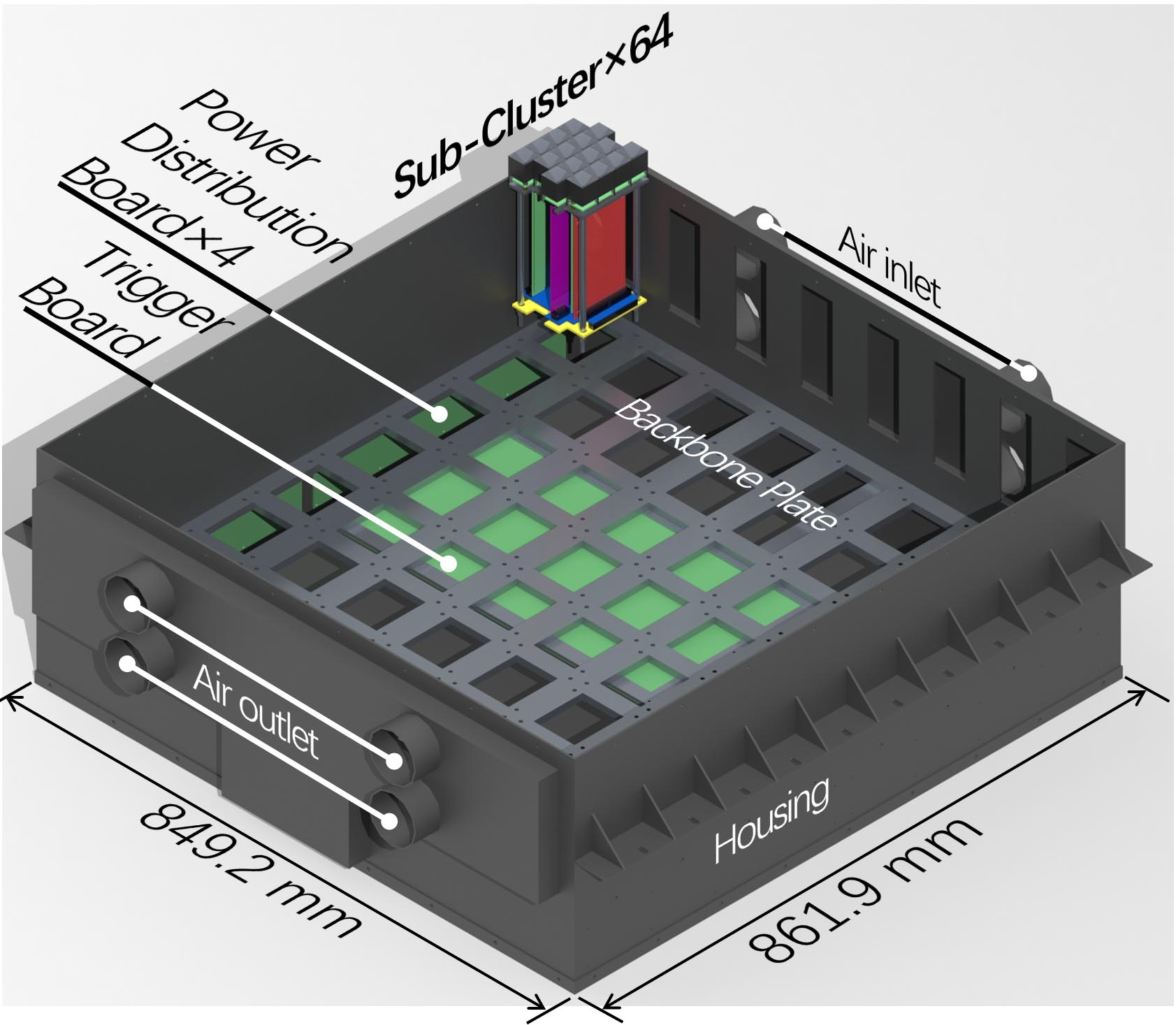}\label{fig:camera_struct}}
	\subfigure[]{\includegraphics[width=6.5cm]{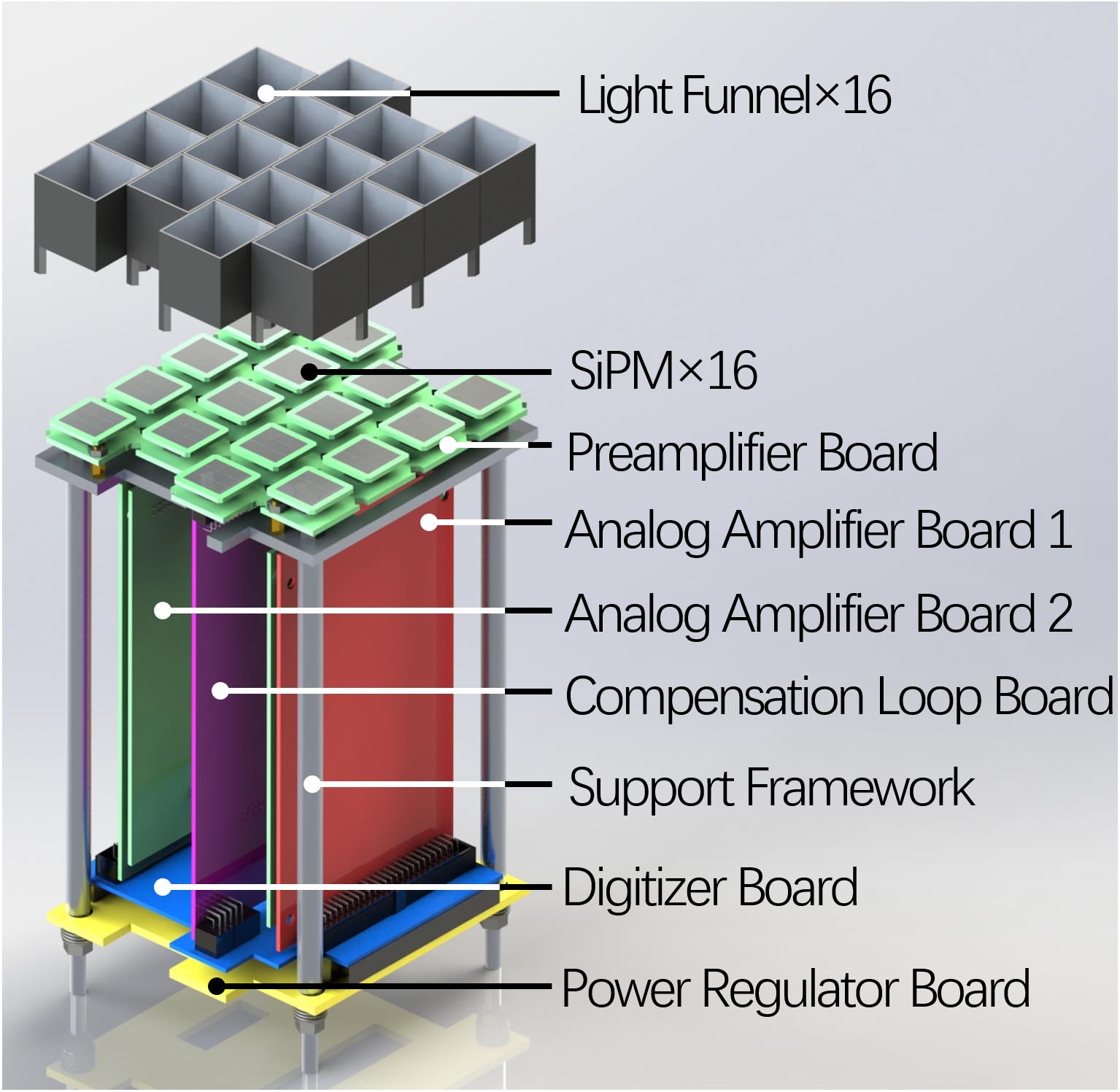}\label{fig:subc_struct}}
	\caption{The structure of the WFCTA SiPM camera (a) and its sub-cluster (b).}
	\label{fig:cam_sc_struct}
\end{figure}

A SiPM camera of WFCTA consists of 64 sub-clusters, a trigger board, four power distribution boards, an aluminium alloy backbone plate, and a steel housing. Each sub-cluster is composed of 16 light funnels, 16 SiPMs, a preamplifier board~\cite{bibaiyang1}, two analog amplifier boards~\cite{xionghao1}, a digitizer board \cite{zhangjinlong1}, a bias voltage and temperature compensation loop board, a power regulator board and an aluminium alloy support frame. The camera is cooled by an air-cooling system. Fig.~\ref{fig:cam_sc_struct} illustrates the configurations of the camera and its sub-cluster.

The WFCTA electronics amplifies and digitizes the signals, and acts as the trigger system and the Ethernet interface. The front-end electronics has two stages. In the first stage, the SiPM current is converted into a voltage signal by a 3~$\Omega$ resistor, and then amplified in a low noise linear preamplifier~\cite{bibaiyang1} (OPA846, Texas Instruments) with a design gain of 10. In the second stage, the preamplifier output is fed in parallel into two linear amplifiers~\cite{xionghao1} (AD8012, Analog Devices) with different gains to cope with the large dynamic range. The first one, with a design gain of 14.1 (high gain), is used for signals with less than 800 p.e., while the second one, with a design gain of 0.64 (low gain), is used to cover the rest of the dynamic range. The width of the amplified signals is stretched to at least 80 ns to avoid sampling aliasing effects.

A 50 MHz, 12 bit analog-to-digital converter (ADC, AD 9249, Analog Devices) digitizes the amplified signals with a resolution of 0.5~mV/ADC-count, and feeds them to a field programmable gate array (FPGA, XC6SLX100T-FGG676, Xilinx) discriminating between signals and noise according to a given threshold. A single-pixel trigger is issued when the signal exceeds the threshold. In this case, the trigger is sent to the trigger board, while the data are buffered in a pipeline and readout only if a telescope trigger is issued by the trigger board. A telescope trigger is issued when the number of triggered pixels exceeds a threshold number and their distribution has a circular (for Cherenkov mode) or linear (for fluorescence mode) pattern in a time window of 1.92~$\mu$s. When a telescope trigger is issued, the FPGA frames the waveform data of triggered pixels into a set of TCP/IP packets, stamps them with a White Rabbit time mark~\cite{whiterabbit_lip,duqiang01} and sends them to the data center. In the data center, a final shower-trigger decision is made based on triggers from other WFCTA telescopes and/or other LHAASO detectors. The waveform data of each pixel is composed of 28 points, and each point is the sum of 4 consecutive original sampling points with a sampling interval of 20 ns. The signal charge is obtained by integrating the waveform with a time window of 320 ns (one point before the pulse peak and two points after the peak) and subtracting the baseline of the waveform.
\subsection{SiPMs}
\subsubsection{General Specifications}
\begin{figure}[hbtp]
	\centering
	\includegraphics[width=12cm]{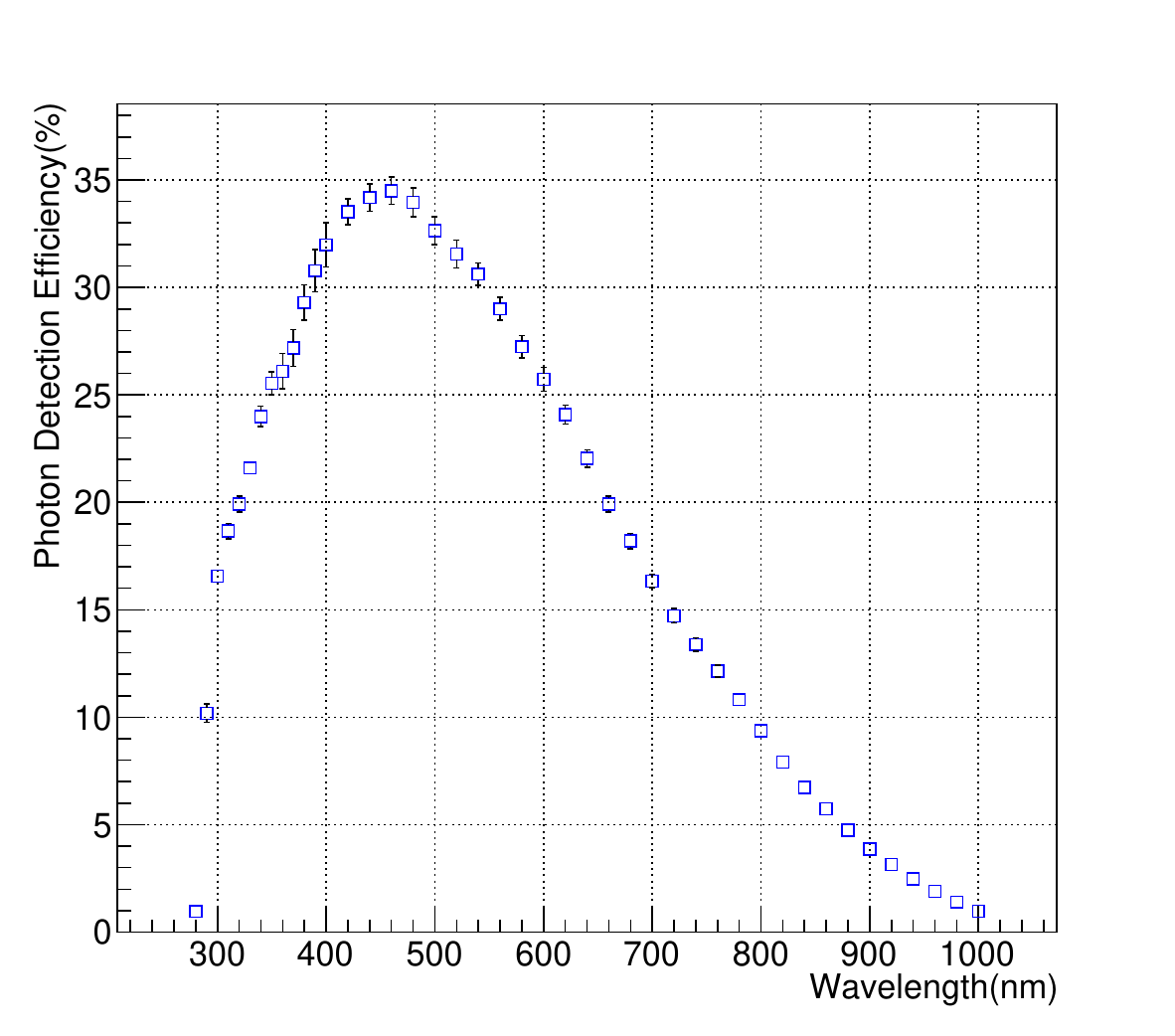}
	\caption{$PDE$ vs. wavelength characteristic of five Hamamatsu S14466 SiPM samples. The blue square is the mean value of $PDE$ of the samples, and the error bar is the RMS value of $PDE$ of the samples. The data were provided by Hamamatsu.}
	\label{fig:pdevslamda}
\end{figure}

A SiPM consists of a large number of Geiger-mode avalanche photodiodes (G-APDs)~\cite{renker1}, each of which is referred to as a cell. The SiPMs used in WFCTA are Hamamatsu S14466. The arrangement of the channels and the temperature sensor of the SiPMs were customized to comply with WFCTA requirements. This type of SiPM has a monolithic array of $3\times3$ channels. Each channel has 39,936 cells, and a size of 5 mm $\times$5 mm. Such a large number of cells are employed because of the dynamic range requirement (10~p.e.-32,000~p.e.). In our design, the output signals of nine channels are summed and read out together. The main specifications of the SiPM are listed in Table \ref{table:s14466sipm}. The photon detection efficiency ($PDE$) vs. wavelength characteristic of the SiPM is shown in Fig.~\ref{fig:pdevslamda}.
\begin{table}[!hbtp]
	\begin{center}
		\begin{threeparttable}
			\caption{Specifications of the WFCTA SiPM (Hamamatsu S14466).}\label{table:s14466sipm}
			\begin{tabular}{p{15em}p{5em}p{3em}}
				\toprule
				Parameters & Value & Unit\\
				\midrule
				Sensitive area & 15$\times$15 & mm$^{2}$ \\
				Number of cells & 359,424 & cell \\
				Number of channels & 3$\times$3 & channel \\
				Cell size & 25$\times$25 & $\mu$m$^{2}$ \\
				Fill factor & 47 & \% \\
				Break down voltage\tnote{1}~($V_{bd}$) & 51.1 & V\\
				$V_{bd}$ difference between channels\tnote{2} & 28.6 & mV\\
				Nominal Gain\tnote{3} & 1.1$\times10^{6}$ &  \\
				Junction capacitance~($C_{j}$) & 21 & fF/cell\\
				Quenching resistor value & 300 & k$\Omega$\\
				Time for 99\% recovery & 32 & ns\\
				\bottomrule
			\end{tabular}
			\begin{tablenotes}
				\item [1] The temperature of the SiPM is 20 $^{\circ}$C.
				\item [2] Root mean square (RMS) deviation.
				\item [3] At a nominal overvoltage of 8.5~V.
			\end{tablenotes}
		\end{threeparttable}
	\end{center}
\end{table}

The WFCTA camera is an upgrade of the original design based on PMTs~\cite{zhangshoushan1,gemaomao1}. This upgrade brings benefits and challenges. Conventional PMTs limit the duty cycle of Cherenkov telescopes~\cite{archambault1,ahnen1}, because bright light, such as the moon and stars light, significantly accelerates their aging. On the contrary, SiPMs can extend their duty cycle~\cite{knoetig2013fact,dorner2019fact} given their capability of working under strong light illumination without noticeable aging effect. In addition, SiPMs are insensitive to magnetic fields, which is important for movable telescopes like WFCTA. On the other hand, SiPMs have some nuisances that need to be considered carefully, such as the optical crosstalk, a comparably high dark count rate, and a temperature-dependent working point. Despite these considerations, the nuisances pose some challenges to the use of SiPM. Pioneering experiments such as the First G-APD Cherenkov Telescope (FACT)~\cite{anderhub1} and the Single-mirror Small-size Telescope (SST)~\cite{heller2017} of the Cherenkov Telescope Array (CTA) have proven its feasibility and have provided us with many useful inspirations and application experiences.
\subsubsection{Optical crosstalk, Afterpulse and Dark Count Rate}\label{subsec:xtandap}
The optical crosstalk~\cite{lacaita1993,mirzoyan1,otte2009,piemonte2019} and the afterpulse~\cite{cova1991,acerbi2015} are considered as the correlated noises of incident optical signals, and their effects need to be evaluated quantitatively in the shower reconstruction. The optical crosstalk and afterpulses have been studied by a variety of experimental and theoretical methods~\cite{oide2007,buzhan2009,nagy2014,anderhub2,nagai2019_2}, and the mathematical models~\cite{rech2008,gallego2013,rosado2015,acerbi2019} have been proposed. According to the measurement results of 120 SiPM samples of type S14466, the crosstalk probability is 3.9\%$\pm$0.4\%, while the afterpulse probability is 4.6\%$\pm$1.2\%. The crosstalk probability is the fraction of dark counts with an amplitude of at least 1.5~p.e. The measurement time window of afterpulses is 1~$\mu$s after the initial pulse. The occurrence probability of afterpulses fall exponentially after the initial pulse with time constants far below 1~$\mu$s. The data were provided by Hamamatsu.

In order to understand the optical crosstalk and afterpulses in detail, a Monte Carlo simulation is conducted based on the measured probability and the models reported in~\cite{gallego2013,rosado2015}. Providing that the `8 nearest neighbors' model is valid for the SiPM used, considering crosstalk cascades up to 5, and an average crosstalk probability of 3.9\%, the Monte Carlo simulation results indicate that crosstalk will introduce a relative error of 4\% and a relative RMS fluctuation of 6.5\% when the number of initial fired cells ($N_{f}$) is 10. Due to the saturation effect of crosstalk~\cite{gallego2013}, the relative error introduced by crosstalk decreases slightly for $N_{f}$>2,000, and reduces to 3.2\% for $N_{f}$=32,000. This saturation effect introduces a non-linearity of $-$0.8\% for $N_{f}$=32,000. The time interval between the initial pulse and the afterpulse is a single exponential distribution, the average $1/e$-amplitude time constant of afterpulses measured for five SiPM samples is approximately 69~ns. There will be 99\% of the afterpulse charge in the integration time window of 320~ns. An average afterpulse probability of 4.6\% introduces a relative error of 4.4\% and a relative RMS fluctuation of 6.3\% for $N_{f}$=10. The relative error introduced by afterpulses gradually approaches 4.6\% with increasing $N_{f}$. The relative errors introduced by crosstalk and afterpulses can be considered as a measurement bias, which can be eliminated by the calibration of the camera. The relative RMS fluctuations introduced by crosstalk and afterpulses decrease almost exponentially with increasing $N_{f}$. Although they increase the measurement uncertainty of small signals, the relative uncertainty will decrease with the accumulation of events. In addition, the measurement uncertainty of small signals is dominated by the Poisson fluctuation of the night sky background (NSB), not by crosstalk and afterpulses.

\begin{figure}[hbtp]
	\centering
	\includegraphics[width=12cm]{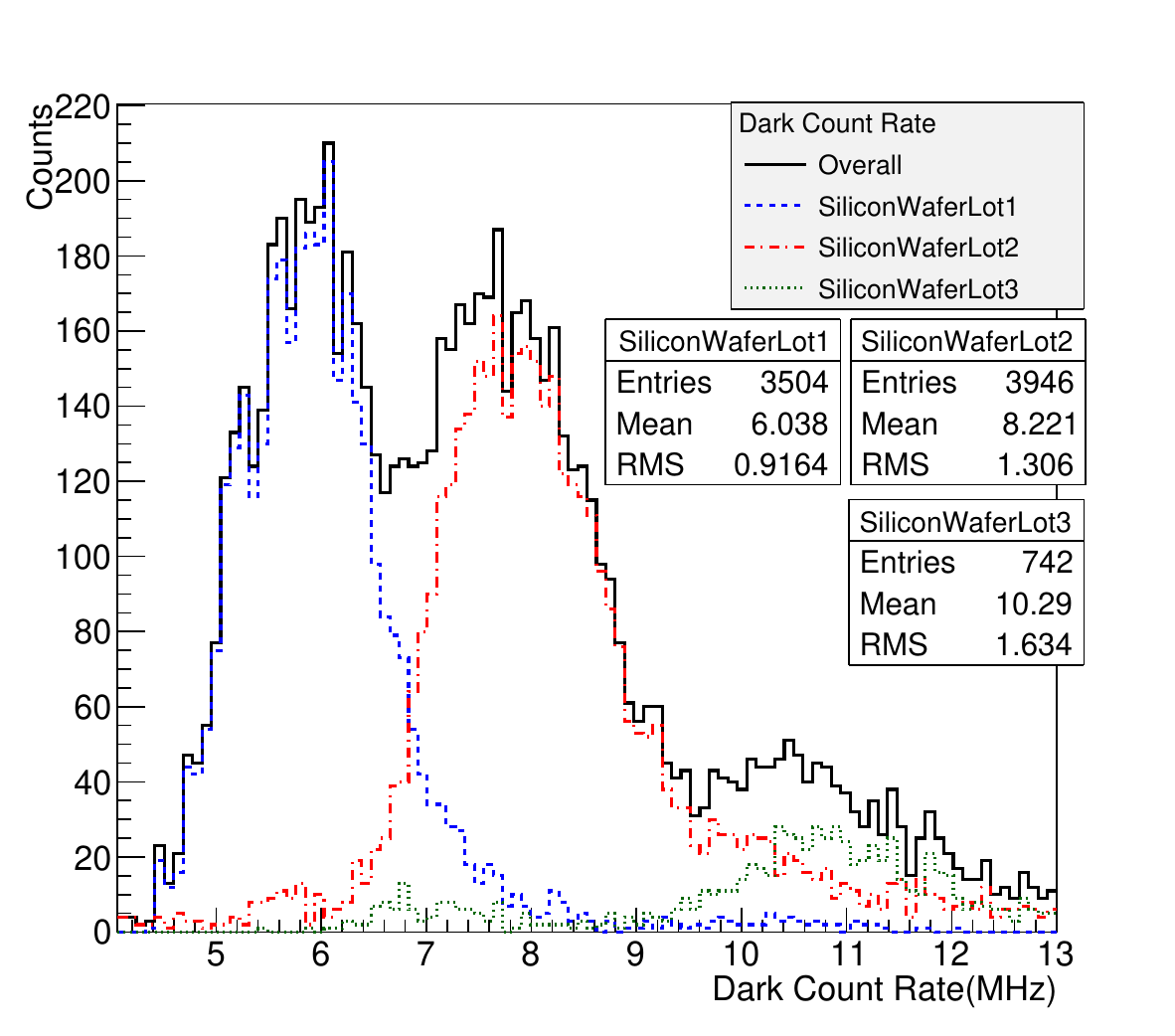}
	\caption{The DCR distribution at 20$^{\circ}$C. DCR is defined per a SiPM. The threshold is 0.5 p.e. The data were provided by Hamamatsu.}
	\label{fig:dcrdis}
\end{figure}

The SiPM generates pulses in dark environments due to thermally generated carriers in the sensitive area. Usually, spuriously triggered breakdowns with an amplitude larger than 0.5~p.e. are considered as dark counts. The number of dark counts per unit time is referred to as the dark count rate (DCR), which increases with increasing operating voltage, temperature and area of SiPMs. For the S14466 SiPM, the DCR changes by about a factor 10 for every 25 $^{\circ}$C change in temperature. Fig.~\ref{fig:dcrdis} shows the DCR distribution for the pixels of the first eight cameras.

It can be seen from Fig.~\ref{fig:dcrdis} that there are three different populations in the distribution. This is mainly because these SiPMs come from different silicon wafer lots. Different wafer lots have subtle differences in their production process that affect the DCR. DCR of the applied SiPM is 4~MHz to 18~MHz, while the count rate induced by the diffuse NSB light is usually greater than 50~MHz. Therefore, in actual observations, DCR becomes a negligible factor compared with NSB.

\subsubsection{Temperature Effects and Compensation Loop}
The basic SiPM input-output characteristic, without considering the effects of non-linearity, crosstalk and afterpulses, can be expressed as
\begin{equation}
	Q=N_{p}\cdot PDE\cdot g\cdot e,
\end{equation}
where $Q$ is the SiPM output charge, $N_{p}$ is the number of incident photons, and $e$ is the charge of the electron. $g$ is the gain of the SiPM, and can be formulated as
\begin{equation}
	g=\frac{C_{j}\cdot V_{ov}}{e},
\end{equation}
where $C_{j}$ is the junction capacitance of the SiPM cell. $V_{ov}$ is the overvoltage, and $V_{ov}=V_{bias}-V_{bd}$, where $V_{bias}$ is the bias voltage provided by the power supply, and $V_{bd}$ is the breakdown voltage of the cell. $PDE$ is affected by $V_{ov}$. Consequently, the output charge $Q$ can be formulated as
\begin{equation}
	Q=N_{p}\cdot PDE\left(V_{ov}\right)\cdot C_{j}\cdot V_{ov}.
\end{equation}
The breakdown voltage $V_{bd}$ increases almost linearly with the temperature increasing~\cite{biland2014,otte2017,klanner2019}. So, providing that $N_{p}$ and $V_{bias}$ are constant, $V_{ov}$ and then $Q$ will decrease with the temperature increasing.

In our design, each SiPM is powered individually by a linear voltage regulator (LR8N8, MICROCHIP) with an adjustment range from 54~V to 64~V and a step size of less than 3 mV. The voltage regulators have a temperature regulation factor of 6~mV/C$^\circ$ on average.

\begin{figure}[hbtp]
	\centering
	\includegraphics[width=12cm]{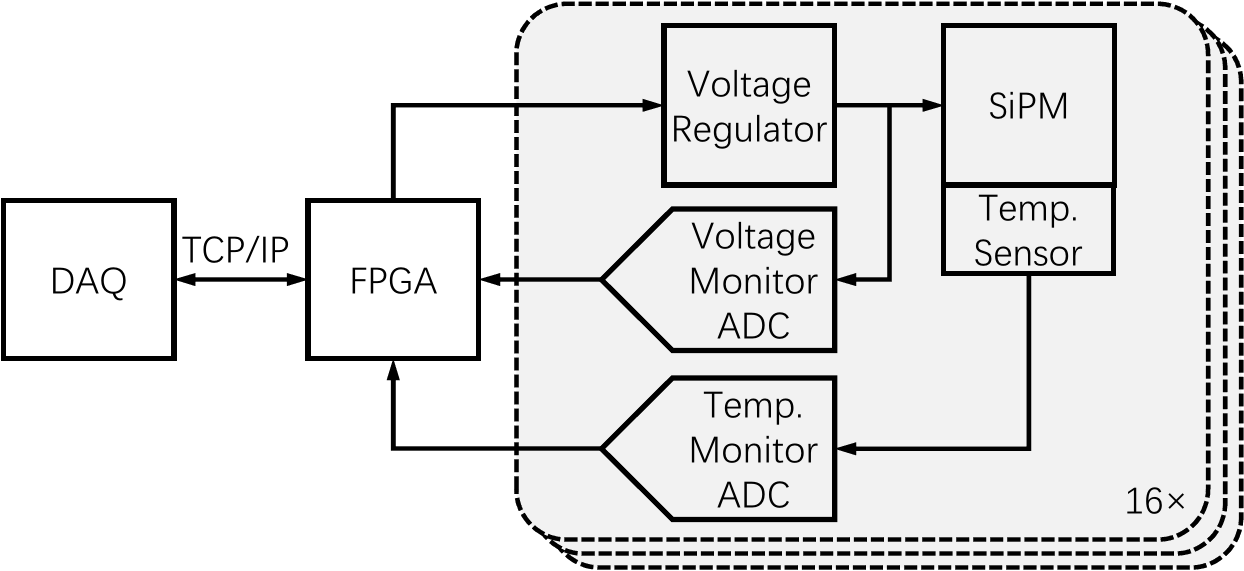}
	\caption{A scheme of the temperature compensation loop.}
	\label{fig:tempcomploop}
\end{figure}

At the LHAASO site, the camera working temperature varies from $-20^{\circ}$C to 25$^{\circ}$C in extreme environmental conditions. To keep the working point stable against temperature variations, a compensation loop has been introduced, which individually adjusts $V_{bias}$ to compensate for the input-output characteristic changes of each pixel due to the temperature variations. A scheme of the temperature compensation loop is shown in Fig.~\ref{fig:tempcomploop}.

\begin{figure}[hbtp]
	\centering
	\includegraphics[width=12cm]{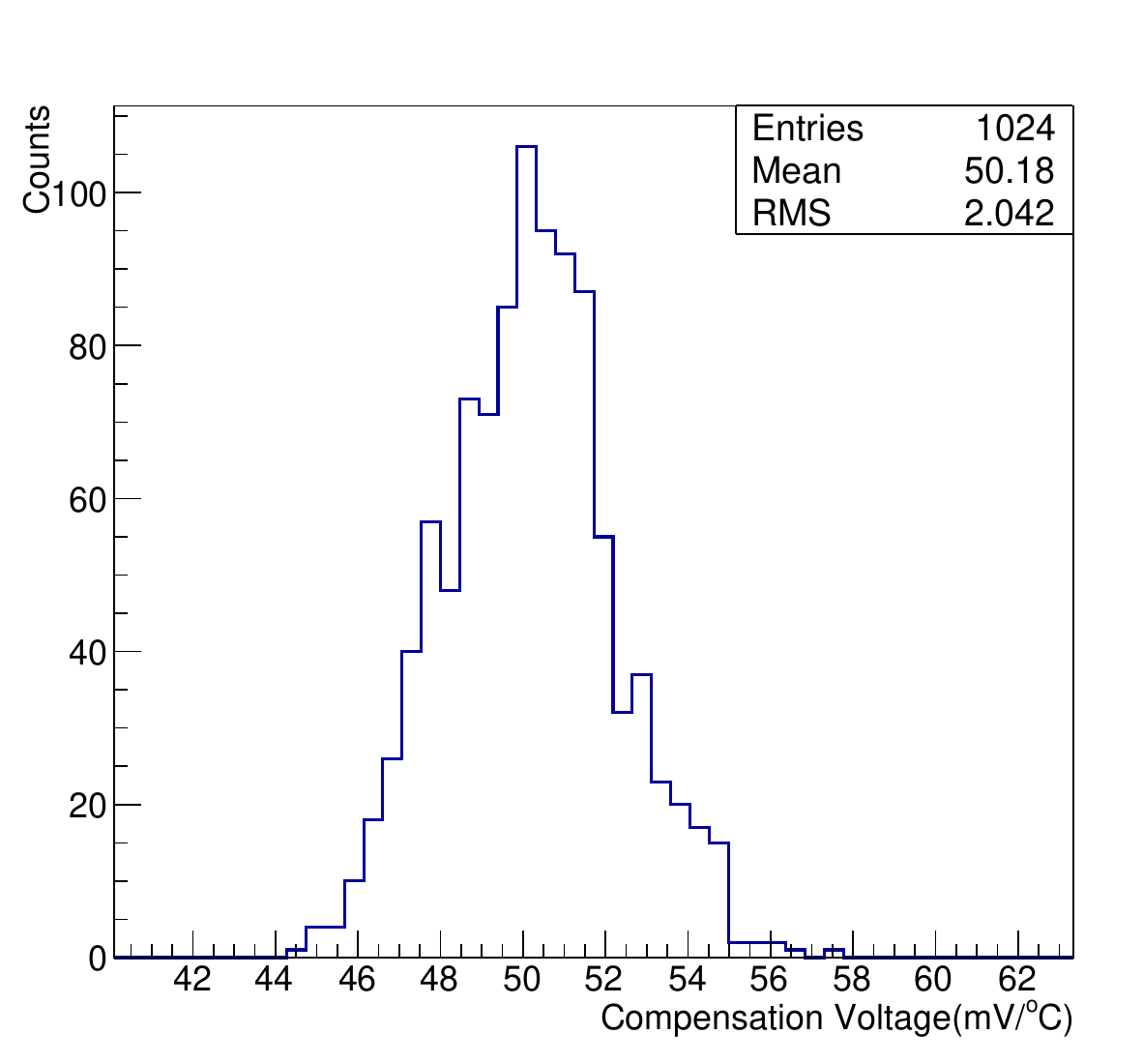}
	\caption{Distribution of the compensation voltage of 1024 pixels in the first camera.}
	\label{fig:tempcompcoeff}
\end{figure}

The compensation applied is based on the on-site measured compensation voltage of each pixel. Fig.~\ref{fig:tempcompcoeff} shows the distribution of the compensation voltage for the 1024 pixels of the first camera. The temperature of the SiPM is monitored individually by a temperature sensor (LM94021, Texas Instruments), which is mounted on its backside. When the power dissipation of the SiPM is very high, there is indeed a difference between the temperature measured by the sensor and the temperature of the SiPM active area. Under normal observation conditions, the whole camera is in thermal equilibrium, and the heat generated by the SiPM itself is negligible. Therefore, the temperature sensor can reflect the temperature of the SiPM active area under normal conditions. The data acquisition (DAQ) program reads the voltage and temperature of the SiPMs every 30 seconds. If the temperature variation of the pixel exceeds $\pm$1~$^{\circ}$C, the compensated $V_{bias}$ is calculated according to its characteristic and the measured temperature. Then the supply voltage is updated to the calculated voltage.

\subsection{Light Funnel}
\begin{figure}[hbtp]
	\centering
	\includegraphics[width=8cm]{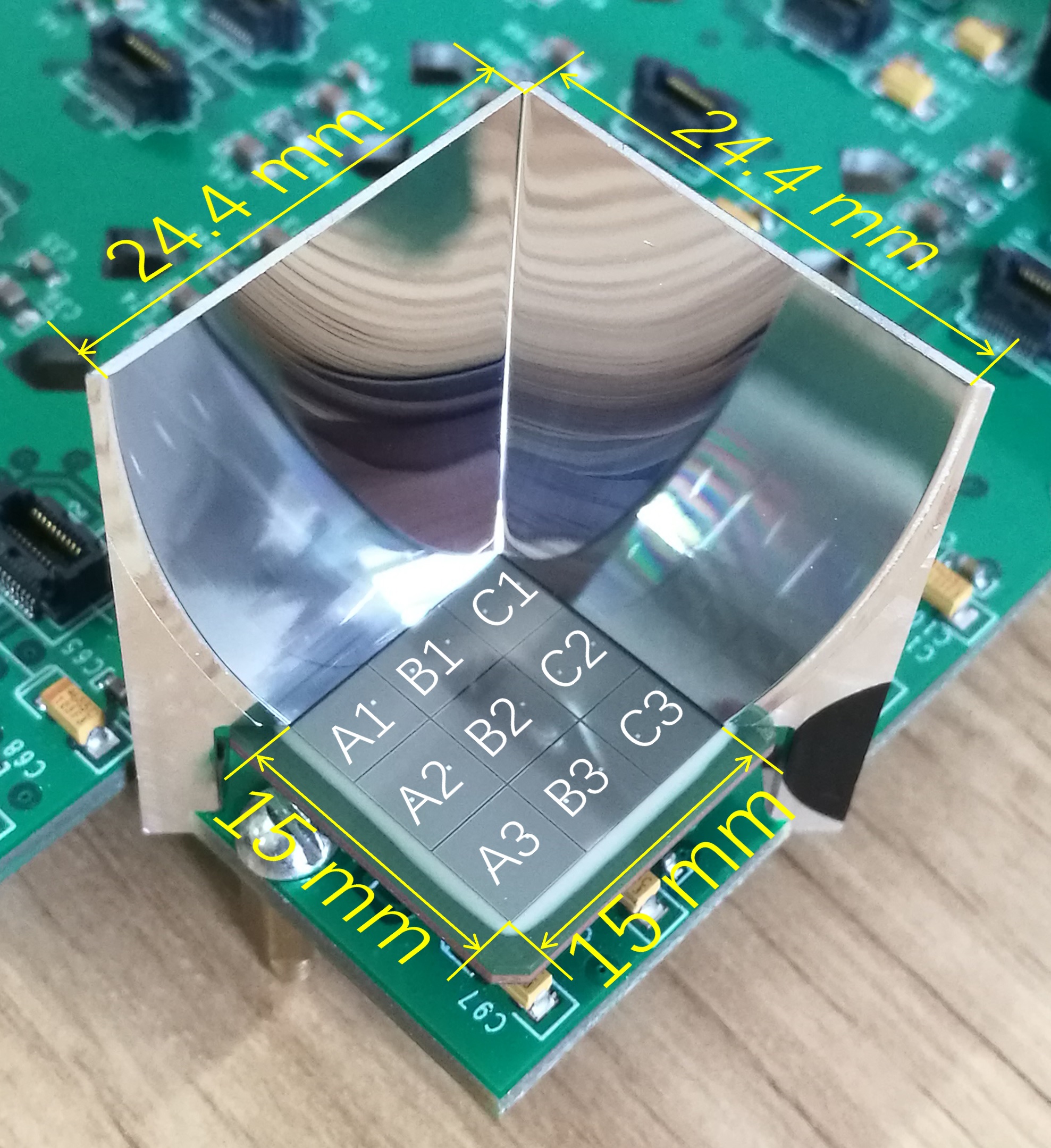}
	\caption{A half-funnel mounted on a SiPM. The marks A1 to C3 are the channel naming convention for the $3\times3$ channels array of the SiPM.}
	\label{fig:halfconeonsipm}
\end{figure}

In order to minimize the dead space between the photosensitive area of SiPMs and to reduce stray light coming from outside the mirror of the telescope, square light funnels are designed to be as close as possible to the ideal Winston cone~\cite{winston1,winston2}, and coupled with the SiPMs. The shape selection of the funnel is constrained by the arrangement of pixels, while it is also a trade-off between the simplicity of implementation and the ideal optical performance. The funnels have an entrance area of 24.4~mm$\times$24.4~mm to match the optical spot size~\cite{liujiali1}, and have an exit area of 15~mm$\times$15~mm to match the size of the SiPM. The sidewall thickness of the funnel is 0.5$\pm0.2$~mm. The clearance between the side walls of the light funnel is 0.2~mm (design value). Therefore, the physical area occupied by each pixel is 25.8~mm $\times$ 25.8~mm. The funnels were designed for a cut-off angle of $35^{\circ}$. The inner surface of the funnels has a roughness of 50 nm, and is coated with an aluminium layer and a dichroic layer (Al+R)~\cite{aguilar201532}. This coating has a reflectivity of about 90\% in the wavelength range from 300 nm to 700 nm, and is provided by Thin Film Physics\footnote{Thin Film Physics AG (TFP), Regensdorf, Switzerland.}. Fig.~\ref{fig:halfconeonsipm} shows how the funnels and the SiPMs are coupled in a pixel.
\begin{figure}[hbtp]
	\centering
	\includegraphics[width=12cm]{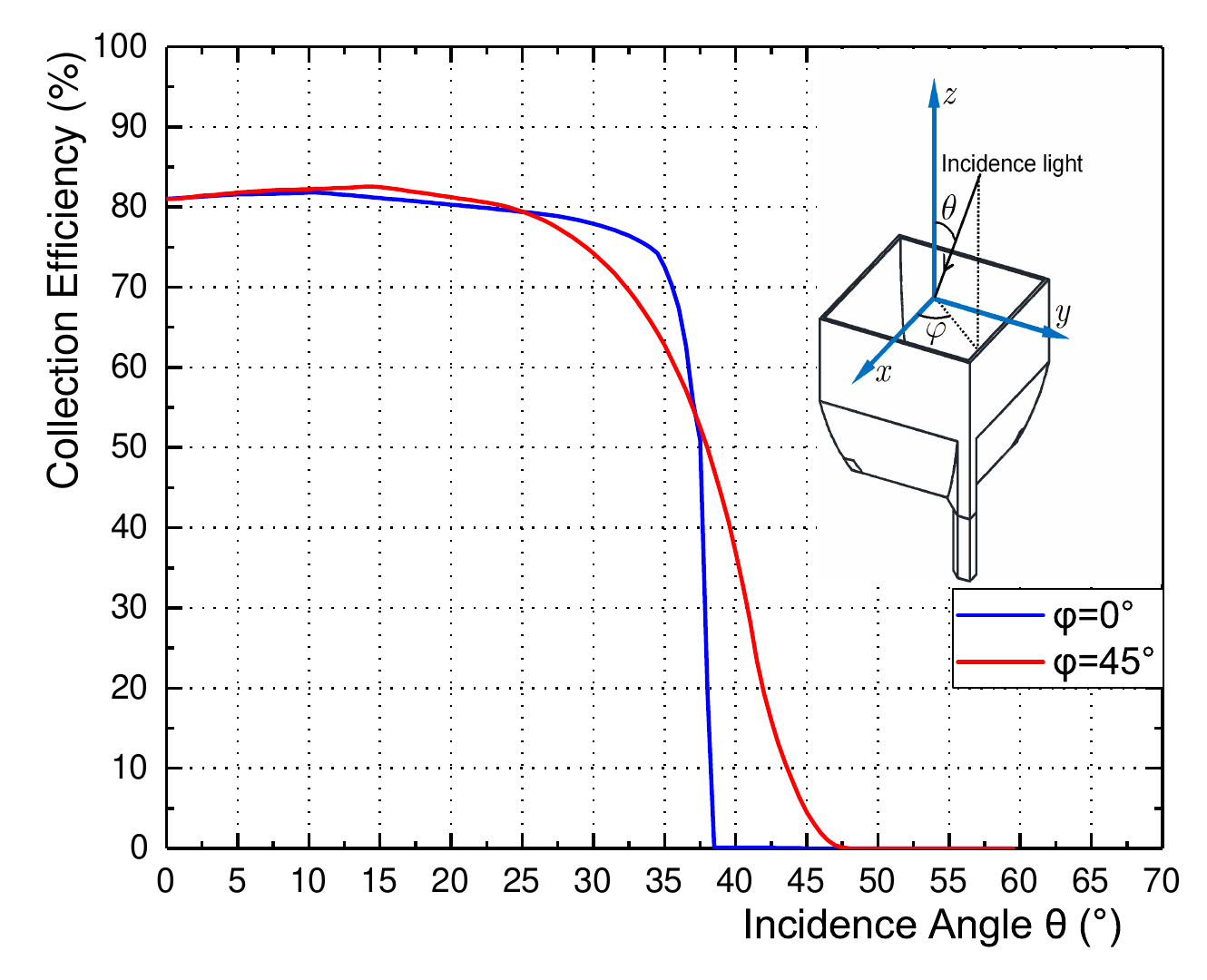}
	\caption{Collection efficiency for parallel beams of light vs. incidence angle obtained from the ray tracing simulation. The blue and the red lines are the collection efficiencies for the incidence azimuth angle of $0^{\circ}$ and $45^{\circ}$, respectively. The sub-graph in the upper right corner of the figure shows the definition of the zenith angle $\theta$ and the azimuth angle $\phi$.}
	\label{fig:coleff}
\end{figure}

The collection efficiency for the parallel beams of light with different incidence angles is shown in Fig.~\ref{fig:coleff}, which is obtained from a ray tracing simulation based on the physical parameters mentioned above, and assuming an ideal surface. The collection efficiency is defined as the ratio of the number of photons exited from the light funnel to the number of photons incident on the pixel (25.8~mm $\times$ 25.8~mm). In actual telescopes, the light reflected by the mirror of the telescope are convergent beams of light. According to the simulation results, the collection efficiency has a maximum of 81\% for vertical light. The clearances between the side walls of the light funnel have a significant impact on this collection efficiency. In order to control the clearances and to avoid stress and deformations, the assembly is done with dedicated jigs, which allow effective constraint on the clearances and the size of sub-clusters.

\section{Laboratory Tests and Assembly}\label{sec:sec3}
\subsection{Test Facilities}
\begin{figure}[hbtp]
	\centering
	\includegraphics[width=8cm]{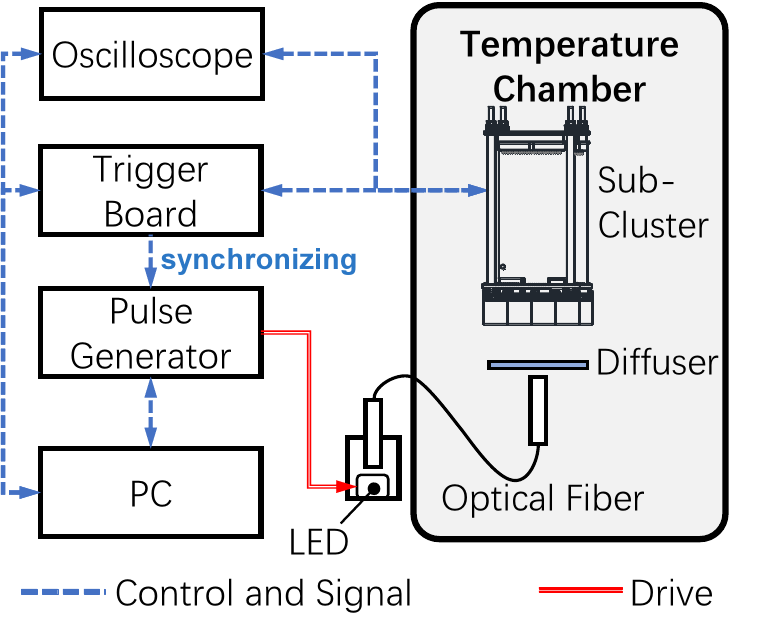}
	\caption{The scheme of Temp-System. The rounded rectangle represents the temperature chamber, which is utilized to provide the temperature environment required for tests. The boxes represent the data acquisition and the control apparatus.}
	\label{fig:tempsys}
\end{figure}

In order to perform a massive test on the performances of SiPMs and sub-clusters, two dedicated facilities have been built. One is called Temp-System, which is used for the temperature dependent characteristics inspection of a small amount of samples (5\% of the total), such as the gain, the response vs. voltage, and DCR. Another is called 1D-System, which is used for the massive batch test on the response vs. voltage, the non-linearity and the signal resolution. The scheme of Temp-System is shown in Fig.~\ref{fig:tempsys}. The LED (NSHU 551AE, NICHIA) light source is placed outside the temperature chamber to avoid having the temperature affect the light output. The light signal is guided into the chamber by an optical fiber, and illuminates the tested SiPMs through a diffuser. The LED light source is powered by a pulse generator (BNC 577-8C, Berkeley Nucleonics), which is synchronized with the trigger board and/or an oscilloscope (DPO 5204B, Tektronix). The pulse used has a frequency of 500~Hz and a width of 20~ns. A set of automatic measurement control and data management software tools are developed to accommodate the massive test tasks. The data processing and drawing libraries of ROOT~\cite{root_framework} are linked with the software tools to graphically monitor the test results in real-time.

\begin{figure}[hbtp]
	\centering
	\includegraphics[width=12cm]{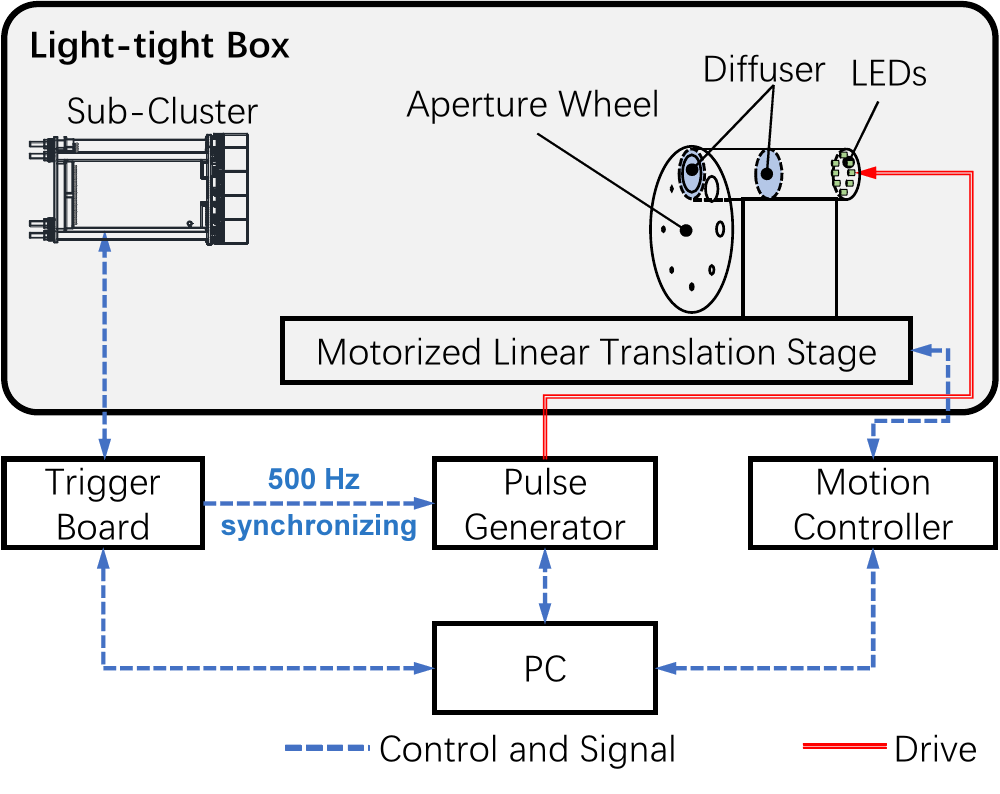}
	\caption{Scheme of the 1D-System. The rounded rectangle represents the light-tight box, which is utilized to provide the dark environment required for tests. The light source is composed of eight LEDs, two diffusers and an exit aperture wheel. The sub-cluster to be tested is illuminated by diffused light whose flux is controlled by the distance from the light source to the SiPM and the exit aperture.}
	\label{fig:1dsys}
\end{figure}

The scheme of the set-up of 1D-System is shown in Fig.~\ref{fig:1dsys}. More detailed information about this system and the test method can be found in Ref.~\cite{gemaomao1}.

\subsection{Response Charge vs. Voltage Characteristic}
\begin{figure}[hbtp]
	\centering
	\includegraphics[width=12cm]{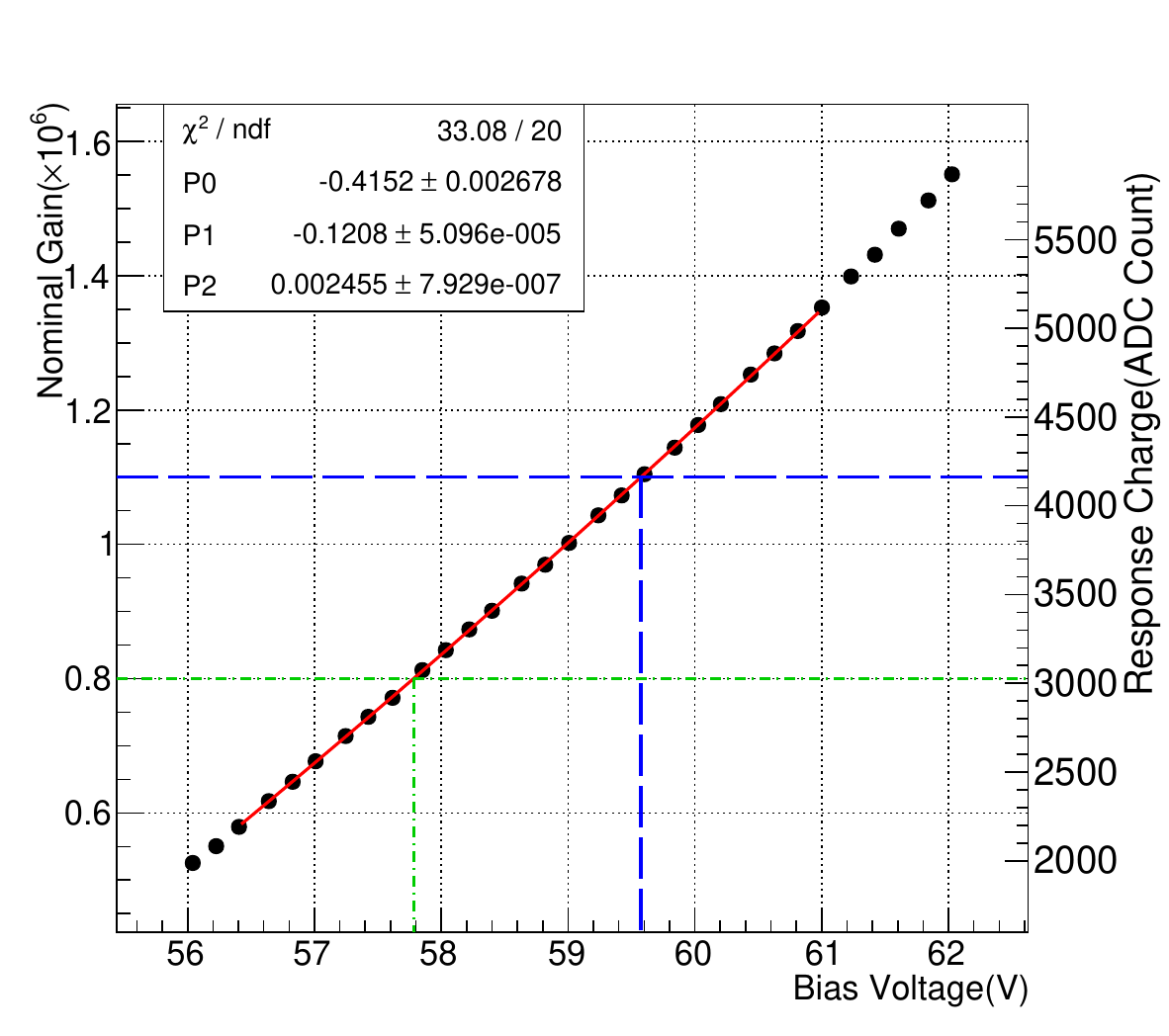}
	\caption{The response charge-voltage characteristic of a SiPM whose serial number is 7517. The SiPM temperature is 20~$^{\circ}$C. The response charge is read from the high gain electronic channel. The red line is a quadratic fitting. The blue dashed line indicates the bias voltage for $g=1.1\times10^{6}$, while the green dash-dotted line indicates the bias voltage for the nominal gain $g=8\times10^{5}$.}
	\label{fig:gvsv}
\end{figure}

\begin{figure}[hbtp]
	\centering
	\includegraphics[width=12cm]{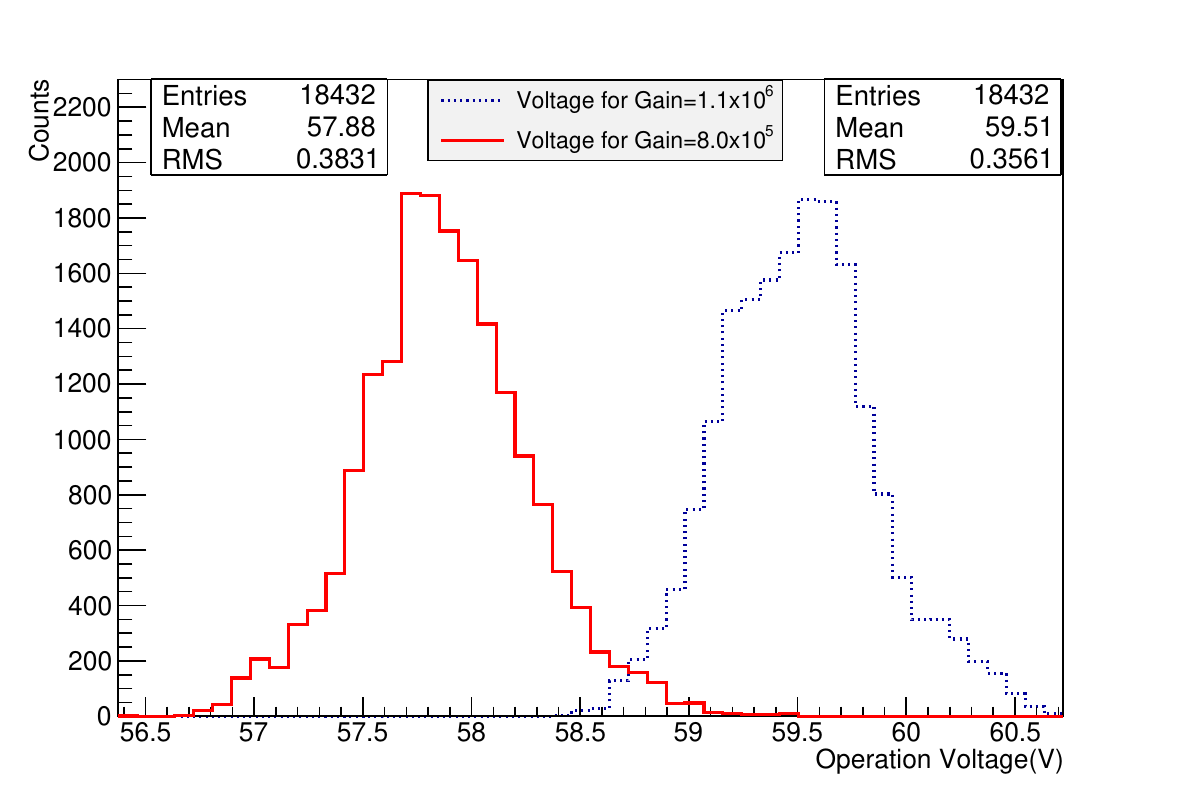}
	\caption{Operation voltage distribution of 18,432 SiPMs in eighteen WFCTA cameras. The dispersion of the operation voltage for gain=$1.1\times10^{6}$ is mainly coming from the process differences in the silicon wafer lots. The standard uncertainty of the measured operation voltage for $g=8\times10^{5}$ is 0.11~V.}
	\label{fig:ophv}
\end{figure}

In different observation phases of WFCTA, the responsivity to the optical signal of each pixel is adjusted individually according to different observation requirements. Therefore, the response charge vs. the bias voltage characteristic of each pixel should be known. Hamamatsu provides the bias voltage ($V_{0}$) of all SiPMs for the gain of $1.1\times10^{6}$ ($g_{0}$). Here, the gain calculated by the following equation is defined as the nominal gain ($g$). 
\begin{equation}
	g=\frac{Q}{Q_{0}}g_{0},
\end{equation}
where $Q$  is the response charge for a certain bias voltage. $Q_{0}$ is the response charge for $V_{0}$. The nominal gain for the first observation phase is $8\times 10^{5}$. The operation voltage for a nominal gain of $8\times 10^{5}$ is given by fitting the response charge-voltage characteristic, as is shown in Fig.~\ref{fig:gvsv}. The SiPM temperature at which this curve is measured is also recorded, so that it can be used for the compensation of the effect of the temperature. The response charge is the integration value of the signal pulse in a time window of 320~ns, in which crosstalk and afterpulses are included. Because the probabilities of them also increase with increasing the bias voltage~\cite{otte2017}, the response charge-voltage curve is slightly upturned. Fig.~\ref{fig:ophv} shows the operation voltage distribution of the SiPMs in eighteen WFCTA cameras for $g=8\times 10^{5}$ and $1.1\times 10^{6}$.

\subsection{Non-linearity and Signal Resolution}\label{subsec:drandsr}
In order to correctly reconstruct physical parameters of CRs, the relation between the flux of photons from a shower (input) and the camera signal charge (output) should be known. To this end, the non-linearity in dynamic range of each pixel is measured and simulated. The measurement setup is shown in Fig.~\ref{fig:1dsys}, and the double-pulse method is used, which can be found in~\cite{bibaiyang1,gemaomao1,hamamatsu_pmt}.

The non-linear behavior of the camera pixels is related to the following three factors: the binomial response of the SiPM to incident photons, the non-uniform photon distribution on the SiPM coming from the pattern induced by the funnel, and the readout electronics non-linearity ($NL_{elec}$).

The binomial response of the SiPM originates from the fact that there is always a certain probability that two or more photons entering the same cell at an interval shorter than the cell recovery time. In this case, the cell is still not fully recharged, and thus the second and succeeding photons will not produce a full amplitude output. The pulse width used in our test is 20~ns, which is less than the SiPM recovery time of 32~ns. And the ratio of the number of fired cells to the number of total cells is lower than 11\%, so the following equation can be used to approximately describe the binomial response for the uniformly illuminated SiPM~\cite{bretz2016,hamamatsu_mppc}
\begin{equation}\label{equ:5}
	N_{f}=N_{0}\left[1-\text{exp}\left(-\frac{N_{p}\cdot PDE}{N_{0}}\right)\right],
\end{equation}
where $N_{f}$ is the number of fired cells, $PDE$ is the photon detection efficiency, $N_{0}$ is the total number of cells, and $N_{p}$ is the number of incident photons. $N_{p}\cdot PDE$ can be considered as the number of photo-electrons ($N_{pe}$) that should be measured without the binomial response. $N_{f}$ is calculated from the signal charge $Q$, and by using the following equation
\begin{equation}\label{equ:6}
	N_{f}=\frac{Q}{g\cdot g_{elec}\cdot e},
\end{equation}
where $g$ is the gain of the SiPM, and adjusted to $8\times10^{5}$ in the test. $g_{elec}$ is the gain of the readout electronics, and $e$ is the charge of the electron. The non-linearity of the uniformly illuminated SiPM ($NL_{SiPM}$) can be calculated by
\begin{equation}\label{equ:7}
	NL_{SiPM}=\frac{N_{f}-N_{pe}}{N_{pe}}=\frac{N_{f}}{N_{0}\ln{\left[N_{0}/(N_{0}-N_{f})\right]}}-1.
\end{equation}

\begin{figure}[hbtp]
	\centering
	\includegraphics[width=12cm]{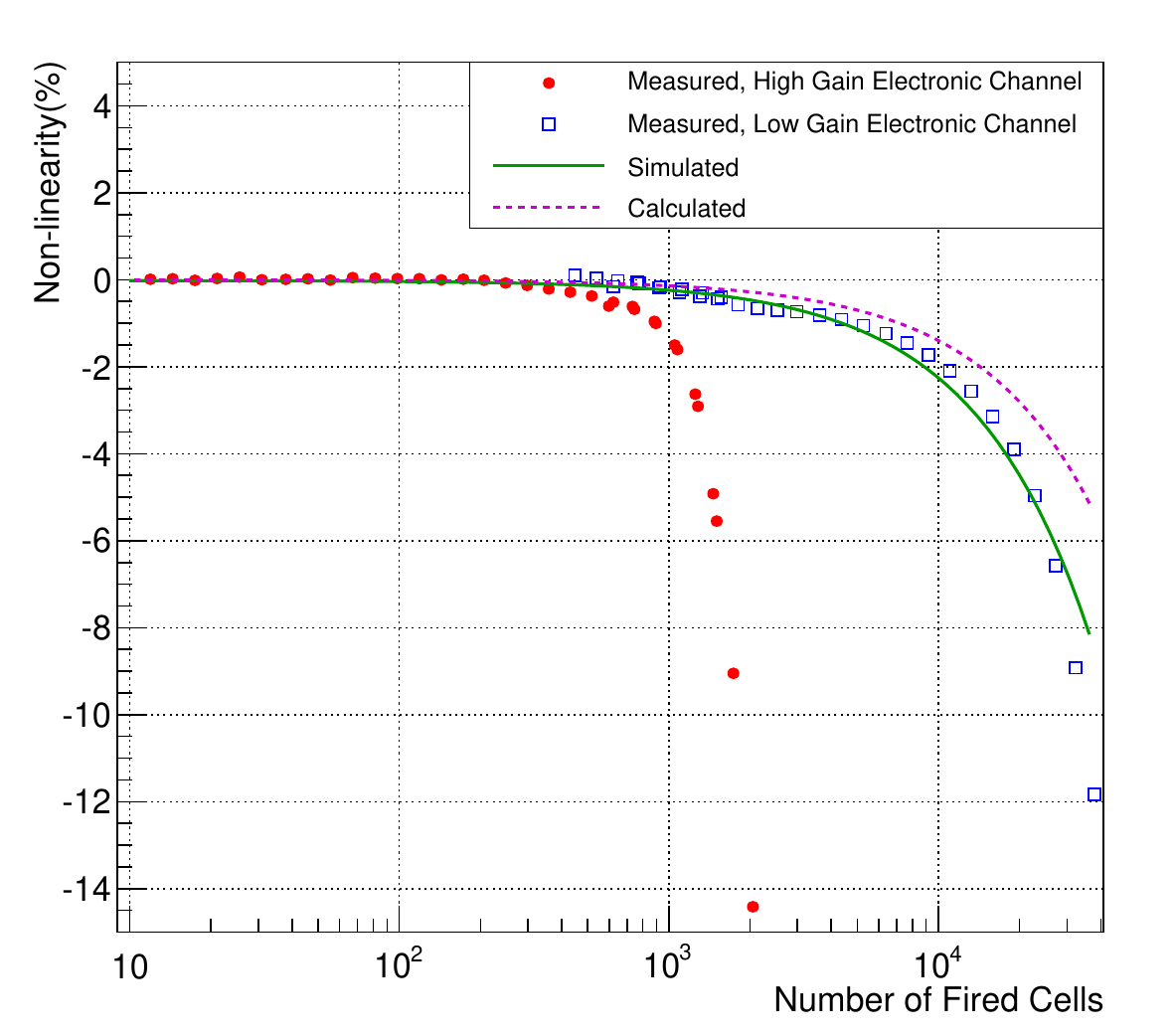}
	\caption{Overall non-linearities of 16,384 pixels of the first sixteen WFCTA cameras. The red circles and the blue hollow squares are the average non-linearities measured by the high gain and the low gain electronic channel of the pixels, respectively. The green line is the non-linearity under the non-uniform illumination which is simulated based on the physical parameters of the light funnel. The purple dashed line represents the non-linearity under the uniform illumination which is calculated from Eq.~\ref{equ:7}.}
	\label{fig:nonlinearity}
\end{figure}

In order to understand the effects of the non-uniform photon distribution, a ray tracing is performed based on the physical parameters of the light funnel and the geometric relationship between the light source and the pixels in the test system. The arrival time distribution of the photons is obtained from the waveform measured by a PMT. A Monte Carlo simulation of the binomial response of SiPMs is then carried out according to the results of ray tracing. In this simulation, the effect of optical crosstalk is taken into account. The effects of afterpulses, DCR, the roughness of the light funnel reflective surfaces, and the reflection of photons on the protective layer surface and the silicon surface are not considered, because these factors have no significant effect on the non-linearity. Fig.~\ref{fig:nonlinearity} shows the non-linearity obtained through the measurement, the calculation and the simulation. The discrepancy between the measurement and the simulation results is less than $\pm$1\% over the dynamic range. We attribute this discrepancy to $NL_{elec}$ and the measurement method. $NL_{elec}$ is required to be less than $\pm$2\% within the dynamic range from 10~p.e to 32,000~p.e. In order to avoid introducing a systematic bias into the reconstruction of showers, an off-line correction is applied according to the results of the telescope simulation.
\begin{figure}[hbtp]
	\centering
	\includegraphics[width=12cm]{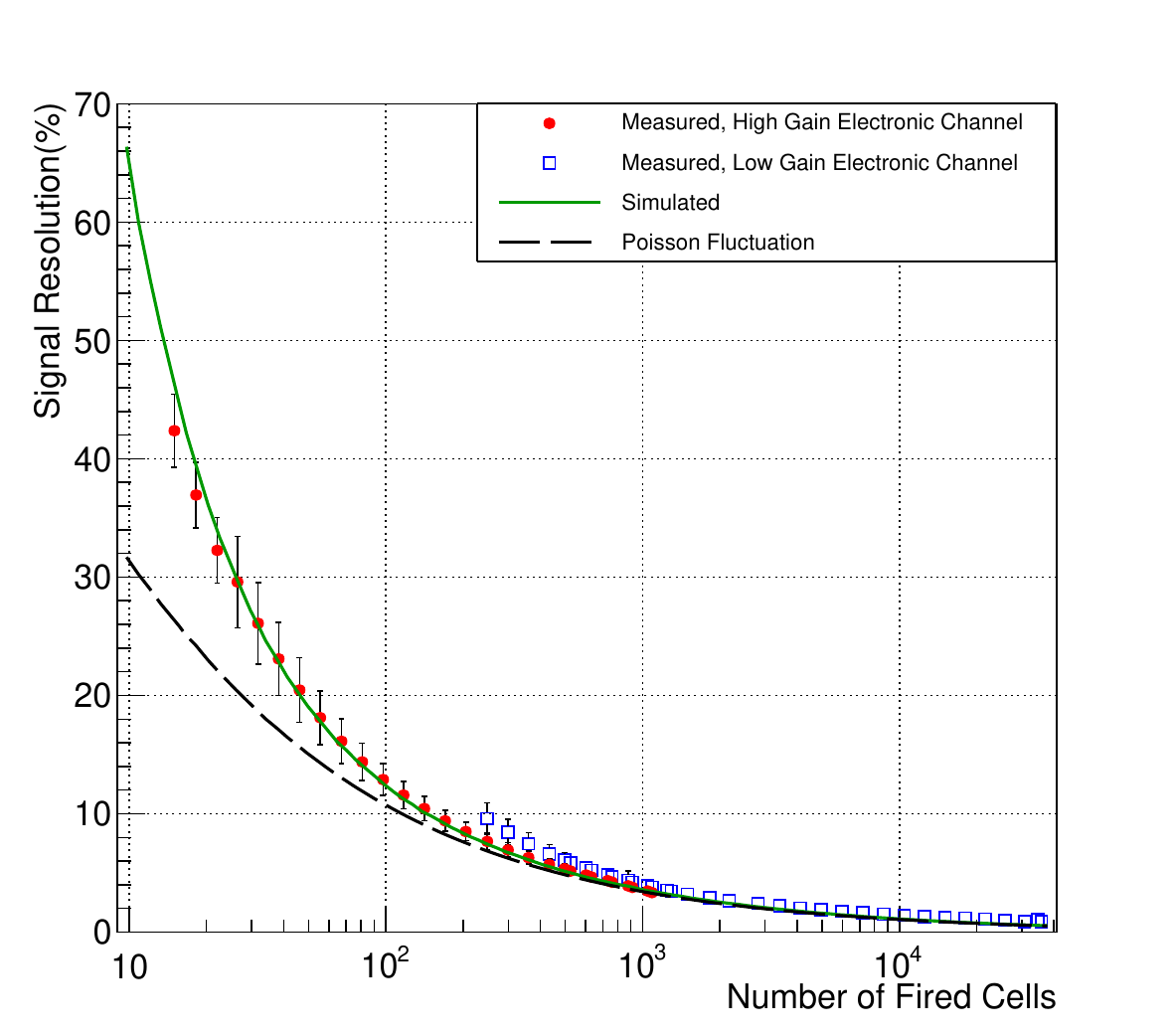}
	\caption{Signal resolutions ($SR$) of 16,384 pixels of the first sixteen WFCTA cameras. The red circles are the average $SR$ of the high gain channels, and the blue hollow squares are the average $SR$ of the low gain channels. The error bar represents the standard deviation of $SR$ distribution of the pixels. The green line is the simulated $SR$. The black dashed line is the Poisson fluctuation calculated from $1/\sqrt{N_{pe}}$.}
	\label{fig:sigres}
\end{figure}

Optical signals will fluctuate after being converted and amplified by SiPMs and readout electronics. This fluctuation is mainly caused by the gain difference between cells\footnote{According to engineers of Hamamatsu, this value is estimated to be less than $\pm$2\%.}, crosstalk and afterpulses of the SiPMs, the noise of the readout electronics, and the Poisson fluctuation during the photo-electric conversion on the SiPMs. We use the signal resolution ($SR$) to quantify the overall effect of these fluctuations. $SR$ is given by 
\begin{equation}
	SR=\frac{\sigma_{Q}}{\overline{Q}},
\end{equation}
where $\sigma_{Q}$ is the standard deviation of signal distribution, and $\overline{Q}$ is the mean value of signals. Fig.~\ref{fig:sigres} shows the measured $SR$, the simulated $SR$, and the expectation of Poisson fluctuation. The models of crosstalk and afterpulses used in simulation have been mentioned in Sec.~\ref{subsec:xtandap}. The number of dark pulses in the integration time window follows the Poisson distribution, and the DCR used is the average value of 7.5~MHz. The fluctuation induced by the noise of the readout electronics is equivalent to (0.65$\pm$0.05)~p.e., which is measured from 1024 readout electronics channels. The differences between the measured $SR$ and the expectation of Poisson fluctuation for $N_{f}$<100 are mainly caused by the noise of the readout electronics and DCR.

\subsection{Assembly Techniques of the Camera}
The assembly process includes four steps: the gluing of the two halves of the funnel, the gluing of the funnels and a preamplifier board (PAB), the assembly of sub-clusters, and the assembly of the camera. To prevent the clearances between the side wall of light funnels from exceeding 0.2~mm, the RMS deviations of length ($L_{C}$) and width ($W_{C}$) of the funnels are required to be less than $\pm$0.1 mm, and the height differences induced by the tilt of funnels to be less than $\pm$0.6 mm. If the assembly processes do not meet the requirements, the collection efficiency will be degraded. Particular care is needed in three of the assembly steps: the correct gluing of the two halves of the funnels to guarantee the same dimensions for all of them, the gluing of the funnels on PAB to ensure the perfect planarity of the funnels' entrance, and the sub-cluster assembly to ensure that all of them have the same height to guarantee the planarity of the whole camera focal plane.

\begin{figure}[hbtp]
	\centering
	\includegraphics[width=8cm]{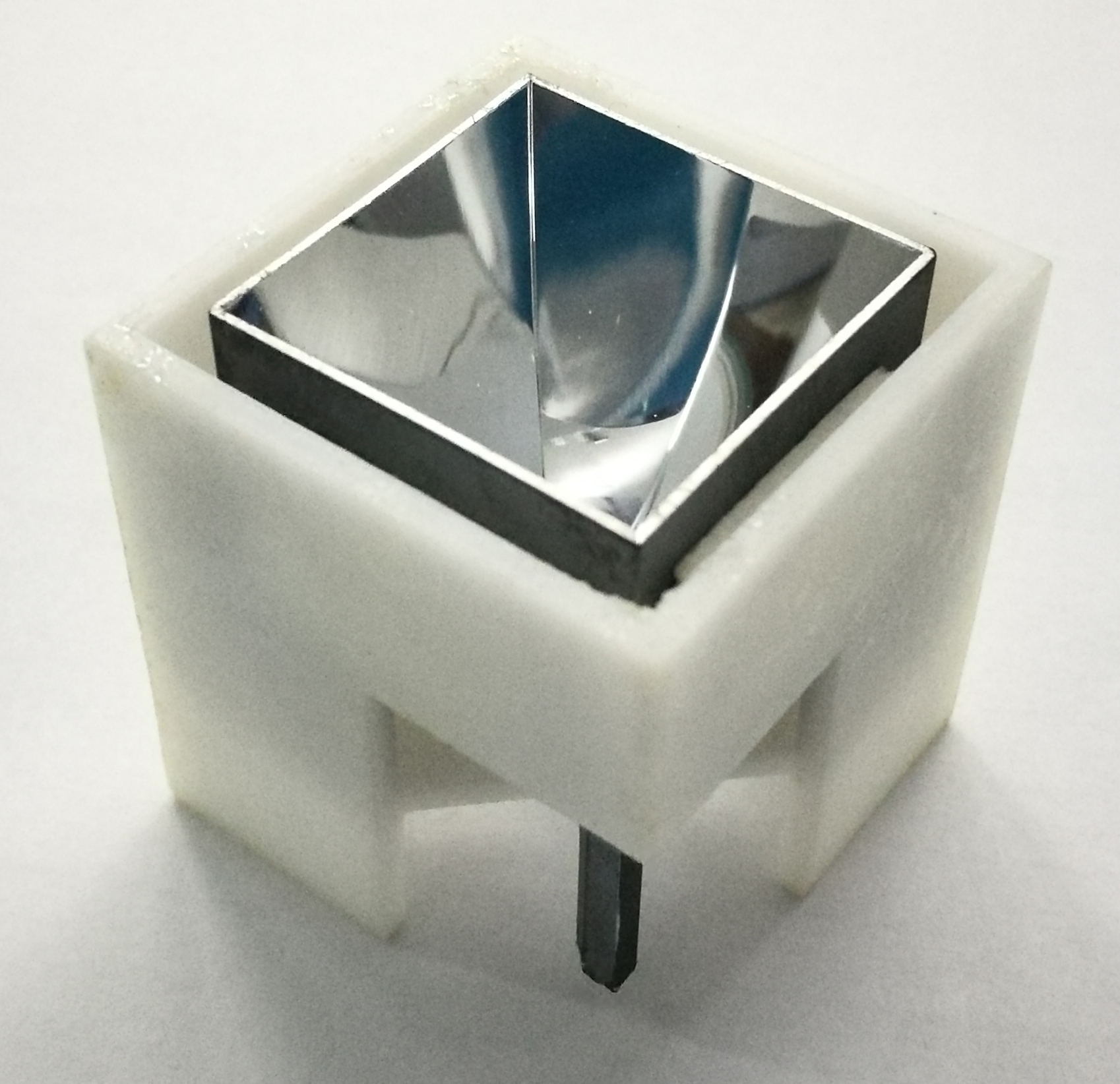}
	\caption{The funnel adhesion jig. The adhesive used is LOCTITE$^{\circledR}$ EA E-20HP.}
	\label{fig:fixture1}
\end{figure}

\begin{figure}[hbtp]
	\centering
	\subfigure[]{\includegraphics[width=6cm]{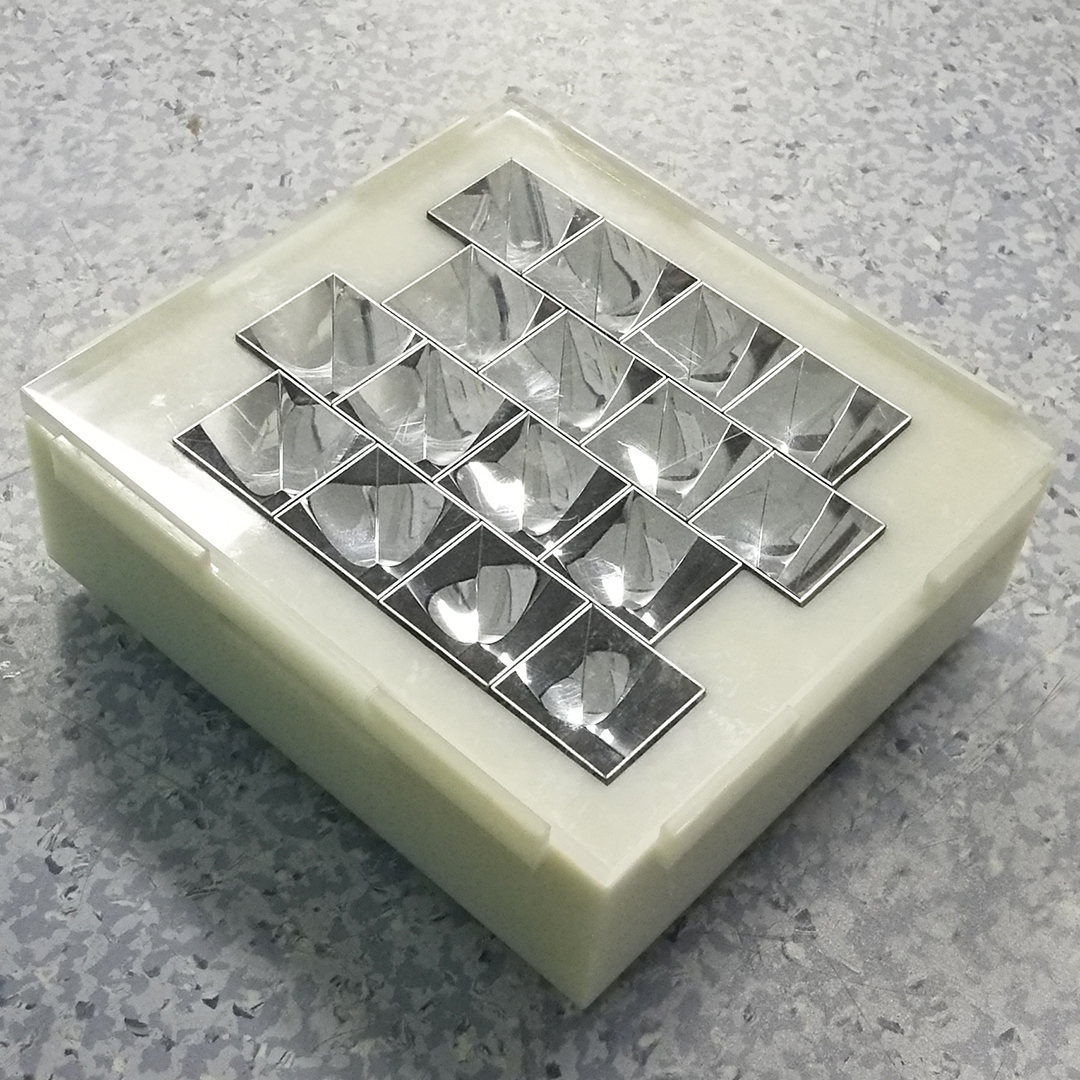}\label{fig:fixture2a}}
	\subfigure[]{\includegraphics[width=6cm]{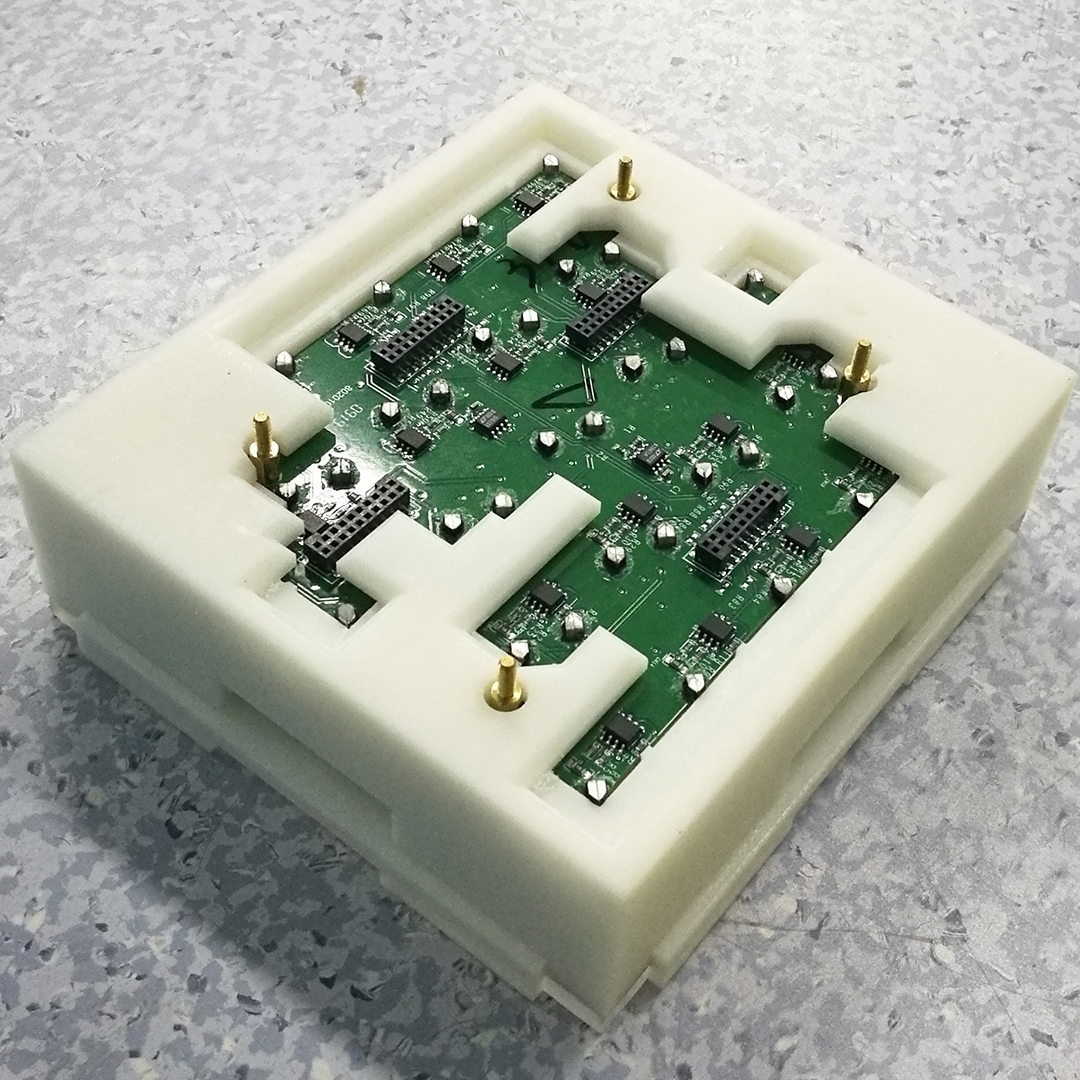}\label{fig:fixture2b}}
	\caption{The funnels-PAB adhesion jig, viewed from above (a) and below (b). The adhesive used is LOCTITE$^{\circledR}$ E-120HP. The jigs are made of ABS resins called Somos$^{\circledR}$ EvoLVe 128.}
	\label{fig:fixture2}
\end{figure}

\begin{figure}[hbtp]
	\centering
	\subfigure[]{\includegraphics[width=8cm]{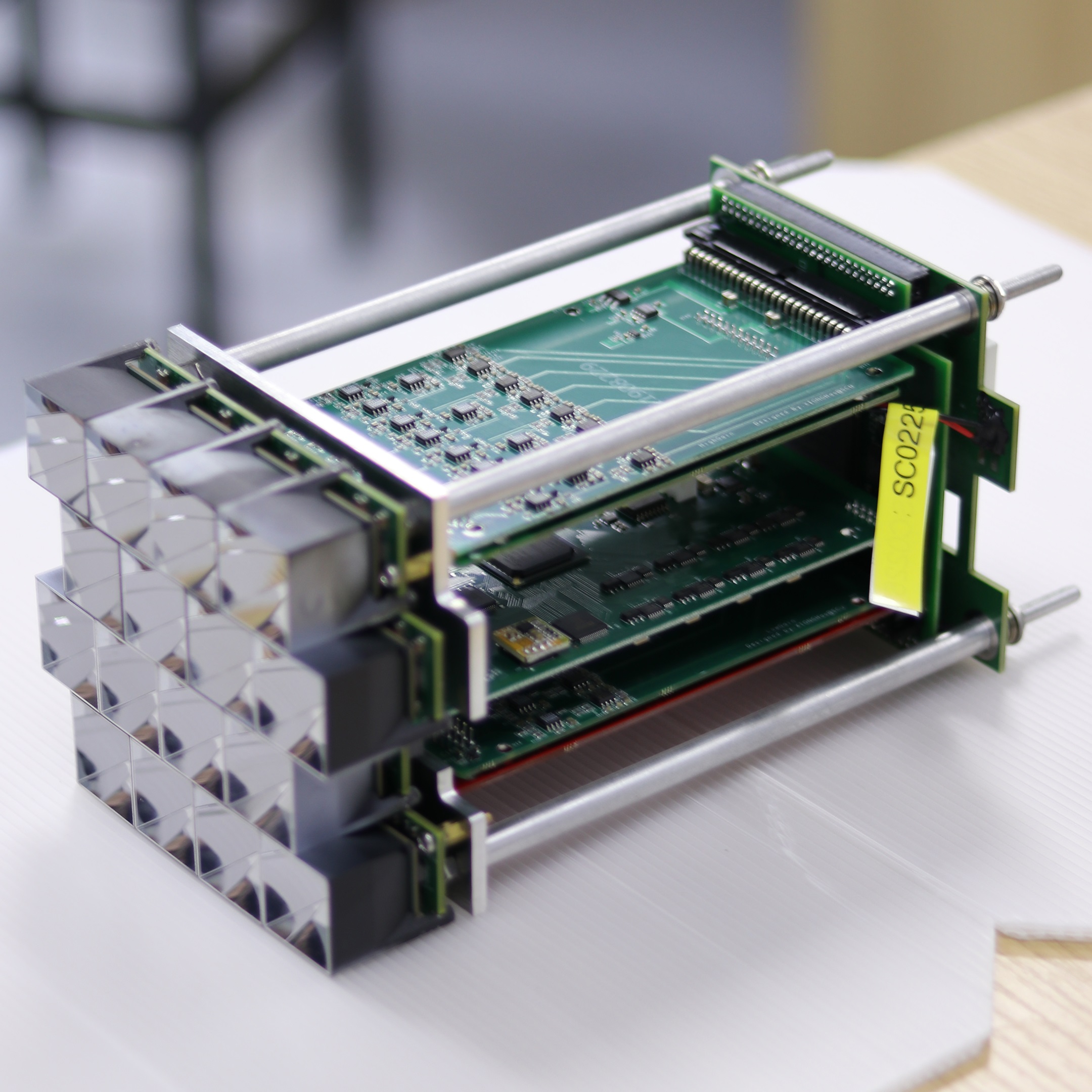}\label{fig:subcluster}}
	\subfigure[]{\includegraphics[width=12cm]{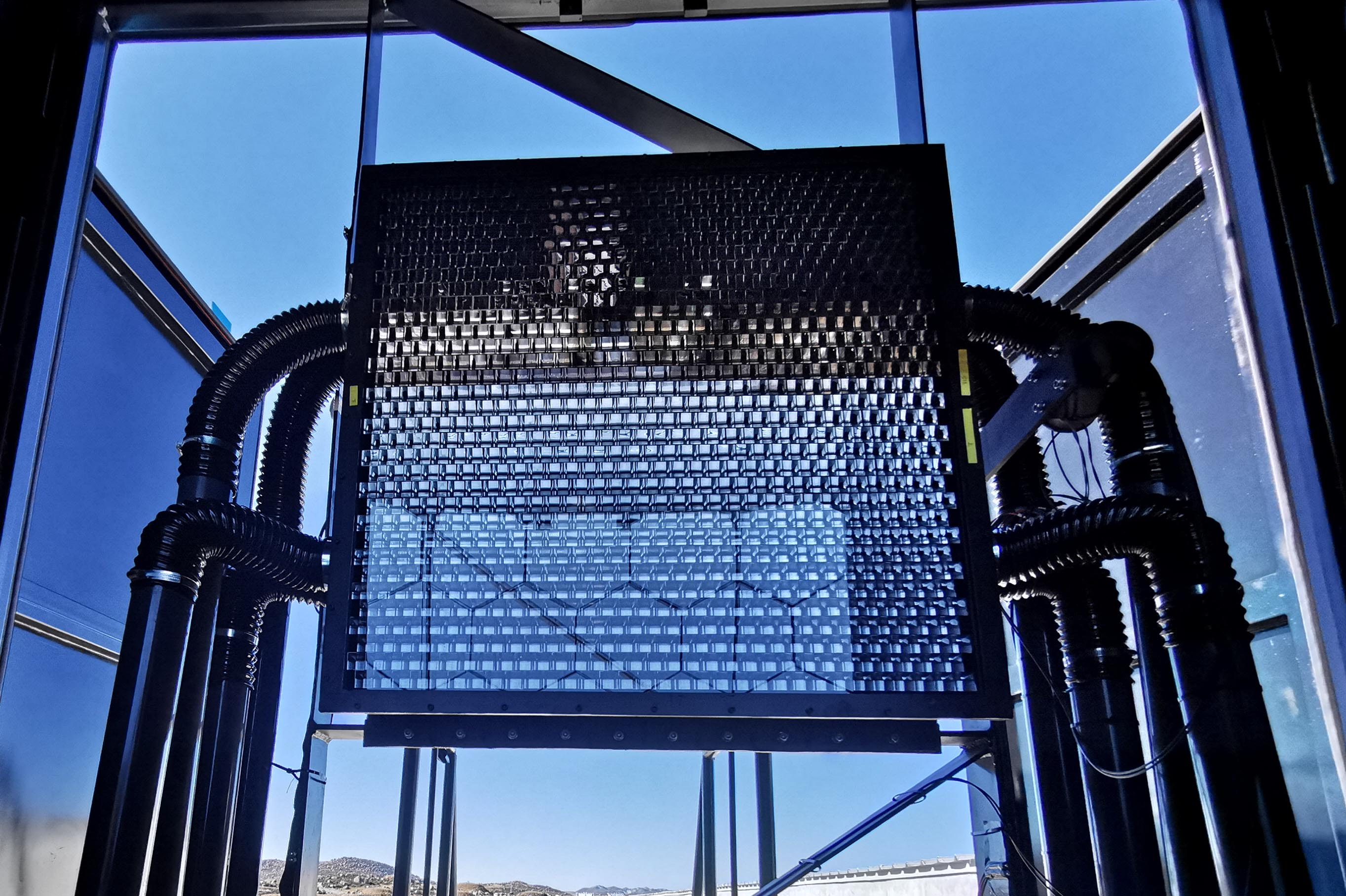}\label{fig:camera}}
	\caption{An assembled sub-cluster (a) and a camera in the telescope (b).}
	\label{fig:sc_camera}
\end{figure}

In order to address the challenges, two kinds of assembling jigs were designed and fabricated by using 3D printing technology. Fig.~\ref{fig:fixture1} shows the funnel gluing jig where funnels are positioned while the glue is drying. The measurement results of 550 randomly selected light funnel samples demonstrate that $L_{C}$ and $W_{C}$ can be kept within $\pm$0.07~mm by this method.
Fig.~\ref{fig:fixture2} shows the funnels-PAB gluing jig, which is similar to a `box' with a removable cover. When the funnels and a PAB are slotted in the jig, the funnels are covered with an acrylic sheet, and then the jig together with the components are turned upside down for gluing. In this way, the entrances of the funnels will be constrained by gravity to lie on the surface of the acrylic sheet. In addition, for every sub-cluster, the height is measured and adjusted until it meets the required tolerance. With this method, the camera focal plane can achieve a flatness of $\pm0.3$ mm. Fig.~\ref{fig:sc_camera} shows a sub-cluster and a camera in the telescope.

\section{On-site Tests and Results} \label{sec:sec4}
\subsection{Running information}
\begin{figure}[hbtp]
	\centering
	\includegraphics[width=12cm]{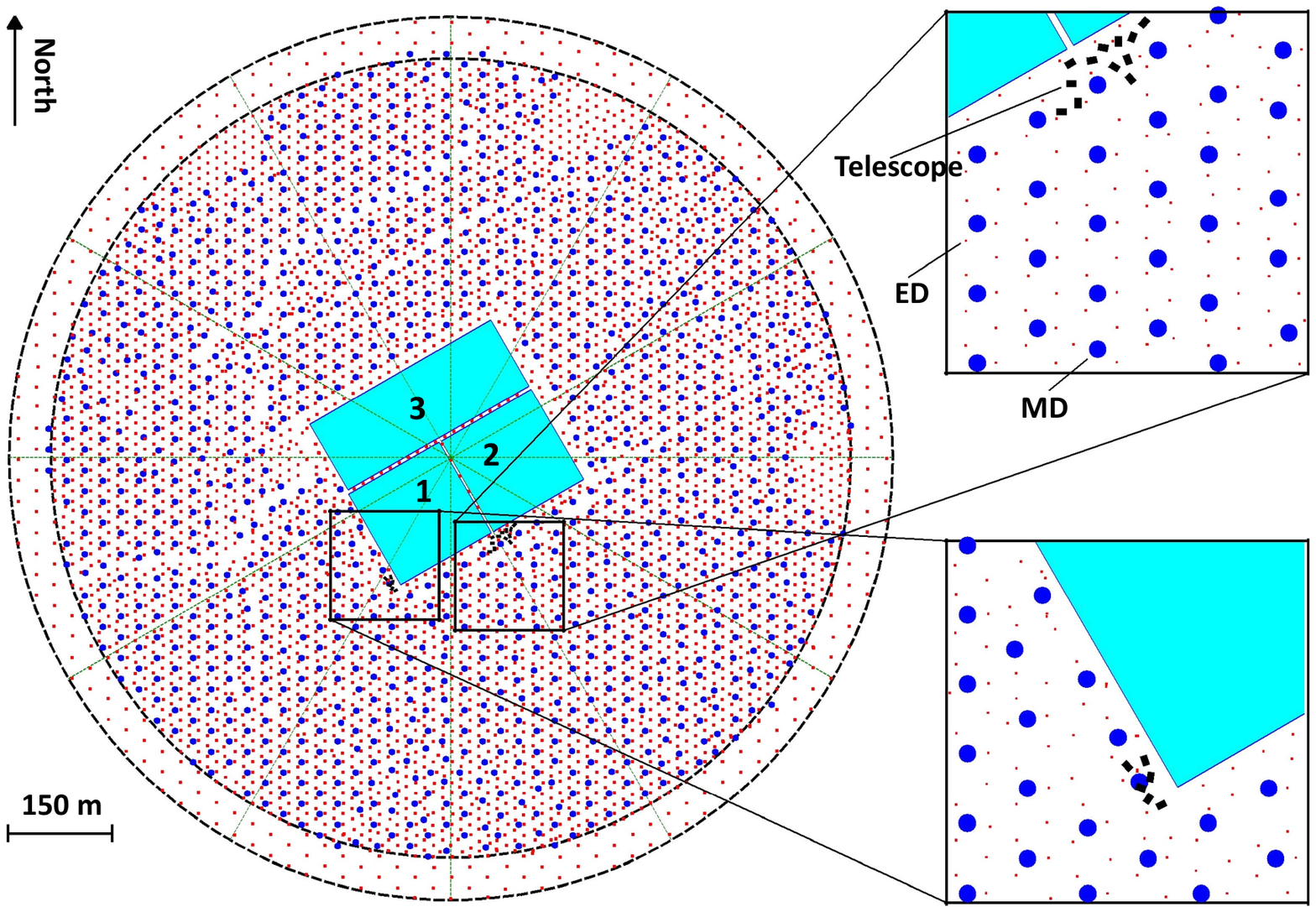}
	\caption{The current layout of eighteen LHAASO-WFCTA telescopes. The small black rectangles indicate the telescopes. Two squares and a rectangle with numbers in the center of LHAASO array indicate three ponds of WCDA, respectively. The square with the number 1 indicates the first pond of WCDA. Six telescopes are located at the southwest corner of the first pond of WCDA and the left twelve telescopes are located at the southeast corner of this pond.}
	\label{fig:WFCTA-array}
\end{figure}

\begin{figure*}[hbtp]
	\centering
	\subfigure[]{\includegraphics[width=12cm]{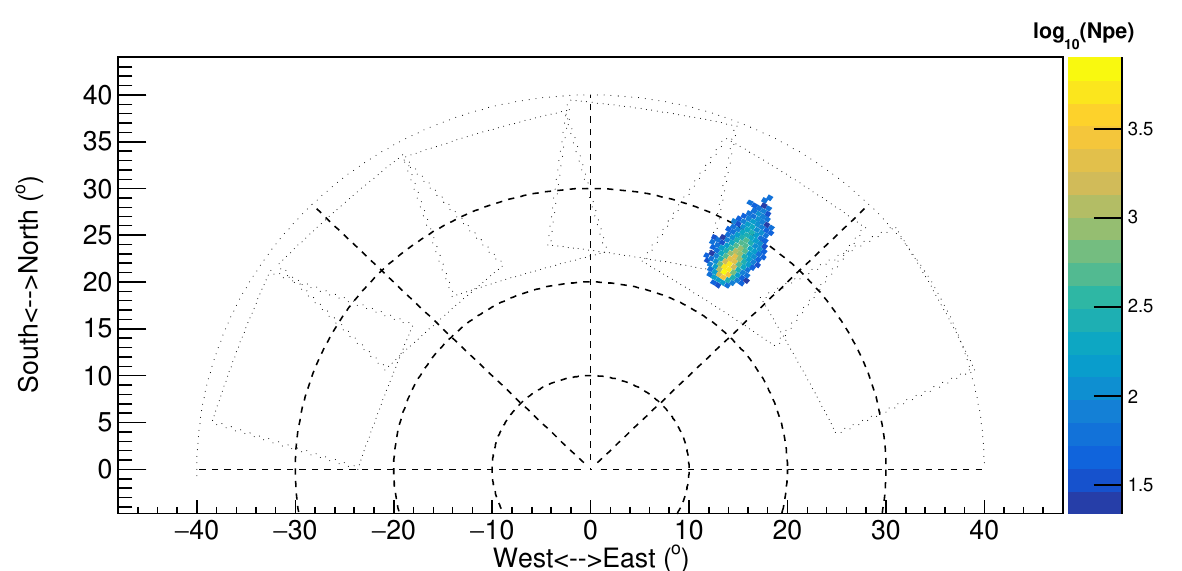} \label{fig:WFCTA-event}}
	\subfigure[]{\includegraphics[width=8cm]{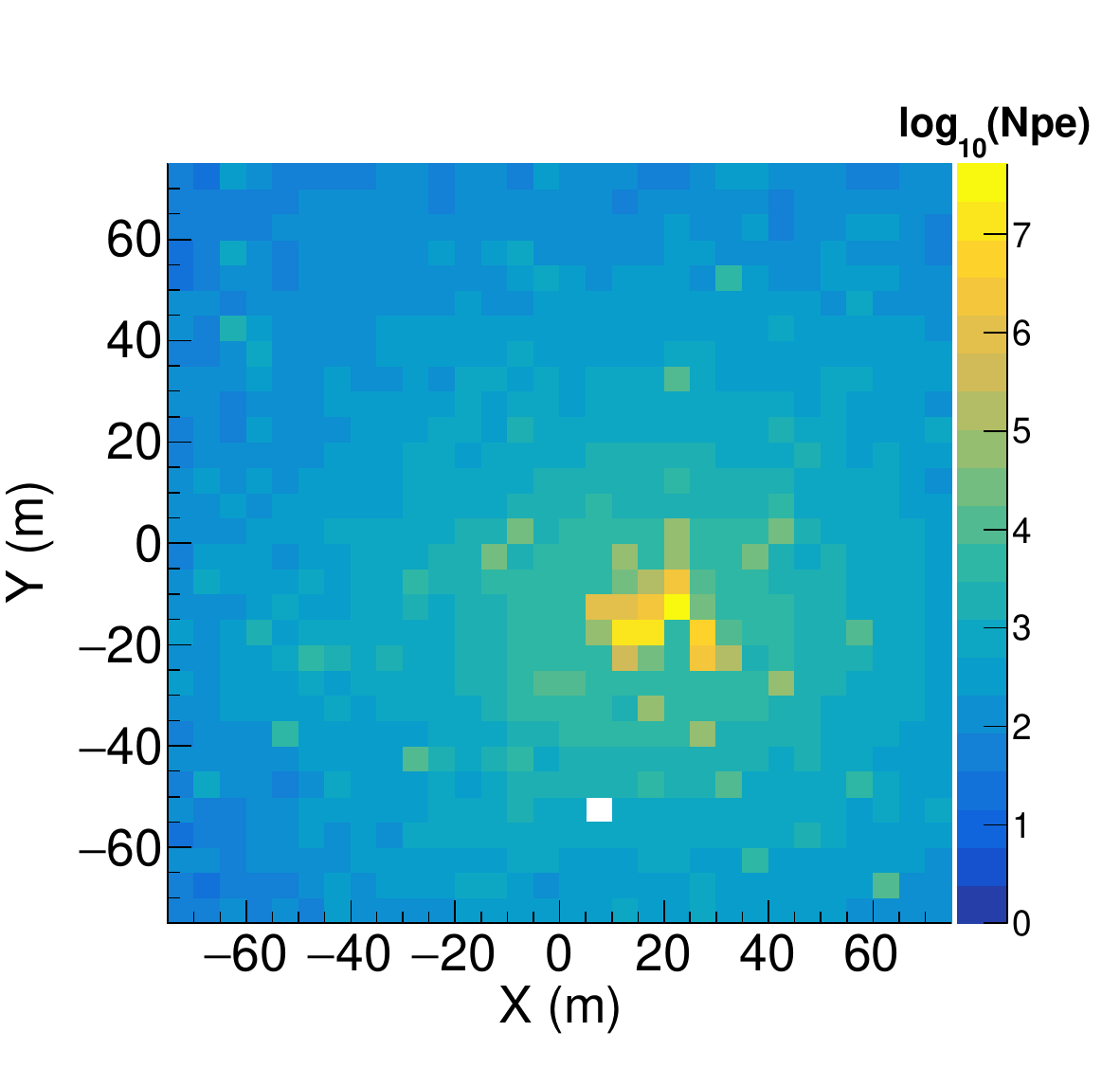} \label{fig:WCDA-event}}
	\caption{A coincidence event as seen by WFCTA (a) and by the first WCDA pond (b). The white pixel in panel (b) is a dead channel. The shower geometry is reconstructed by WCDA. The zenith and azimuth angle are about 24.0$^\circ$ and 57.0$^\circ$, respectively. The shower core position locates at x=15.6~m and y=$-$16.7~m of the first WCDA pond. The perpendicular distance between the shower and the Cherenkov telescope is about 106.6~m. The azimuth of 0$^\circ$ points east and rotates counterclockwise.}
\end{figure*}

The layout of eighteen WFCTA telescopes is shown in Fig.~\ref{fig:WFCTA-array}. The first SiPM-based WFCTA telescope started its operation at the end of January 2019 at the LHAASO site. A total of six telescopes started operation in October 2019. They are located at the southwest corner of the first pond of WCDA. The remaining twelve telescopes are located at the southeast corner of the pond. Two of them started operation in January 2020 and ten in April 2021. 
Events measured by WFCTA are sent to an off-line event filter, which collects them together with events from WCDA or other detectors of LHAASO by using White Rabbit time stamps with a time precision of less than 0.5 ns~\cite{white_rabbit} for time coincidence. More than 50 million coincidence events of the telescopes and the first pond of WCDA were collected by February 2020. One of the coincidence events is shown in Fig.~\ref{fig:WFCTA-event} and Fig.~\ref{fig:WCDA-event}.
\subsection{Performance}
\begin{figure}[hbtp]
	\centering
	\includegraphics[width=12cm]{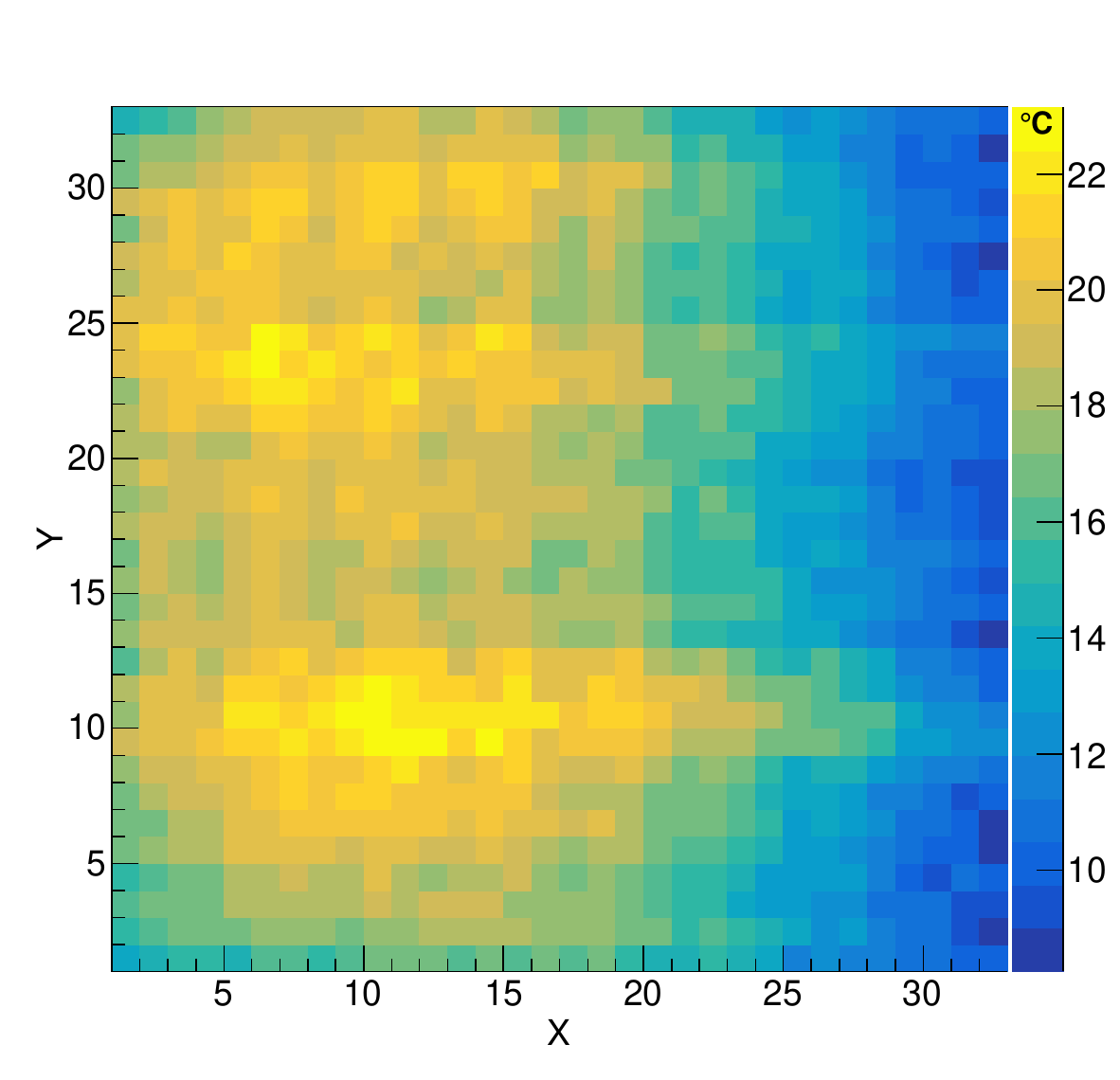}
	\caption{Two-dimensional temperature distribution of 32$\times$32 SiPMs in the camera. The air inlet is on the right side, and the air outlet is on the left side.}
	\label{fig:temperature-distribution}
\end{figure}

The total power consumption of a camera is about 720~W. Each camera is cooled by two 750~W air blowers. The temperature of SiPMs near the air inlet is lower than that of SiPMs near the air outlet. A typical temperature distribution in the camera is shown in Fig.~\ref{fig:temperature-distribution}. 
The gradient between the SiPM with the highest and lowest temperature never exceeds 16~$^\circ$C when two air blowers are running. 
The non-uniformity caused by temperature differences are controlled to be less than 2\% at different ambient temperatures by adjusting the bias voltage of the SiPMs. In addition, the SiPM temperature follows the ambient temperature, and increases during the day and decreases during the night. The typical variation range of the camera average temperature is about 18~$^\circ$C between day and night at the LHAASO site. Bias voltages and temperature compensation loops stabilize the gain of SiPMs to within less than 2\% at different ambient temperatures.

To monitor and calibrate the gain of the SiPM camera, six UV-LED with different wavelength (405~nm, 325~nm, 360~nm, 405~nm, 505~nm, 550~nm) are mounted at the center of the mirror. LEDs and their driving circuit work at a constant temperature of (30.7$\pm0.1$)$~^\circ$C~\cite{yangmingjie1}. An opal glass diffuser is used in front of the LEDs to make a uniform light source.
Every year, all LEDs are calibrated by a calibrated portable probe, which is calibrated by National Institute of Metrology of China.
The camera gain which includes the transmittance of the glass window, the collection efficiency of the light funnel, the SiPM gain and the electronics gain is then absolutely calibrated by the multi-wavelength LEDs. 
One 405~nm LED driven by 35~ns pulse width and 3~Hz frequency pulses is turned on during the telescope observation to monitor the camera gain.

The capability of working in moonlight is the most significant advantage of the SiPM for gamma-ray astronomy and cosmic ray measurement. 
In order to prevent the SiPMs from overheating due to high current under strong light conditions, a bias resistor (R=250 $\Omega$) is connected between the high-voltage (HV) input and the SiPM.
The HV current of the camera is 0.76~A when two shutters of the telescope are closed, and about 0.86~A when the shutters are opened on moonless nights. The maximum output current of the HV power supply is set to 2.5~A, so as to protect the camera from overheating under bright light, such as the moonlight in the telescope FoV during observation or the daylight during the maintenance of the telescope in the day.
At this maximum current, corresponding to about 1.7~mA per SiPM, the heat generated by the internal resistance of the SiPM causes its temperature to rise by 6~$^\circ$C above normal conditions. 
In the maximum current limiting output state, the HV power supply works in constant current mode and the output voltage is not constant.
Although the camera can not be damaged in this condition, the HV power supply is required to be turned off according to the present telescope operation strategy.
Therefore, if the angular distance between the edge of the telescope FoV and the moon is less than 1~$^\circ$, the HV power supply will be turned off automatically to avoid exposing the camera to the direct moonlight, which may lead to the power supply working in the maximum current limiting output state.

The SiPM signals are read out by using a direct current (DC) coupled electronics system. Thus, the baseline amplitude of SiPM output varies with the light intensity of NSB, such as a bright star or the moon passing through the FoV of a pixel in the camera. A higher NSB light intensity generates a higher DC output in the SiPM and then a higher baseline amplitude. The NSB is calculated off-line by the following equation
\begin{equation}
	\text{NSB}=BL_{open}-BL_{close},
\end{equation}
where $BL_{open}$ and $BL_{close}$ stand for the open-shutter and closed-shutter baseline amplitudes, respectively. The amount of moonlight collected by the telescope depends on the moon phase, the atmospheric conditions, the elevation angle of the moon and the angular distance of the moon to the FoV.
\begin{figure}[htp]
	\centering
	\subfigure[]{\includegraphics[width=10cm]{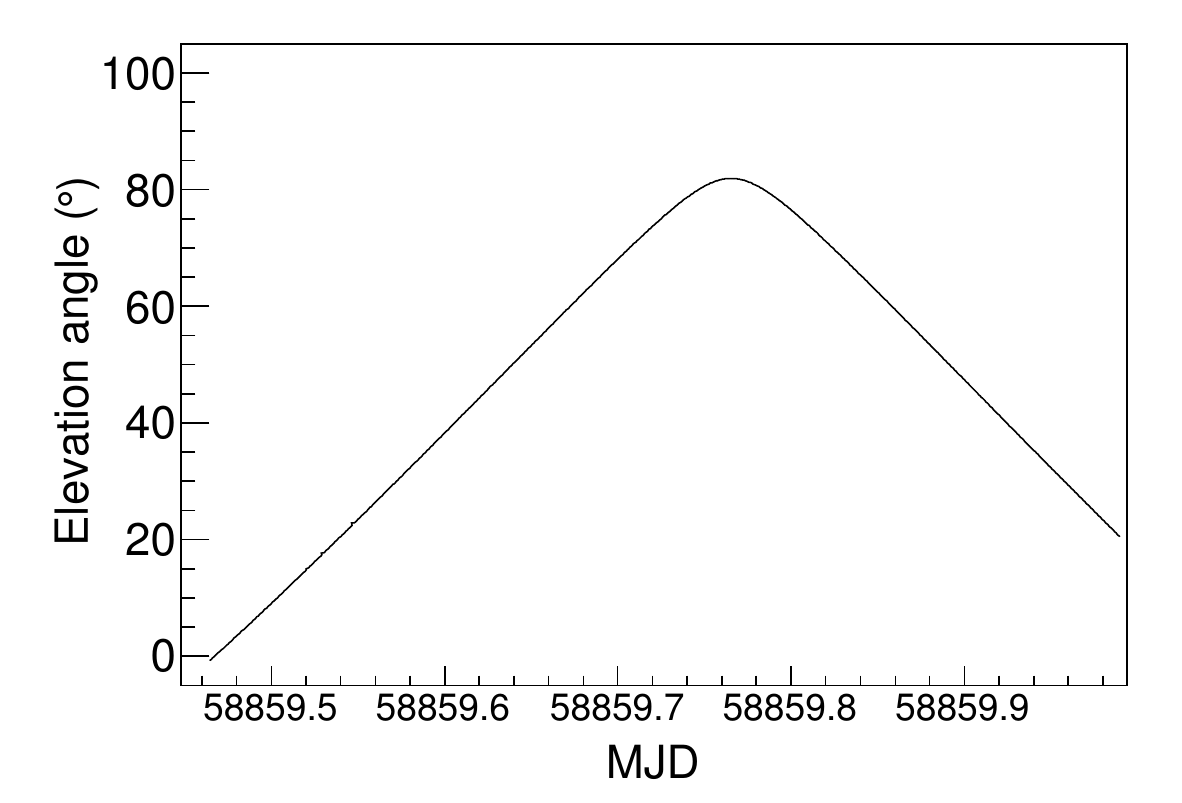} \label{fig:moon-trajectory}}
	\subfigure[]{\includegraphics[width=10cm]{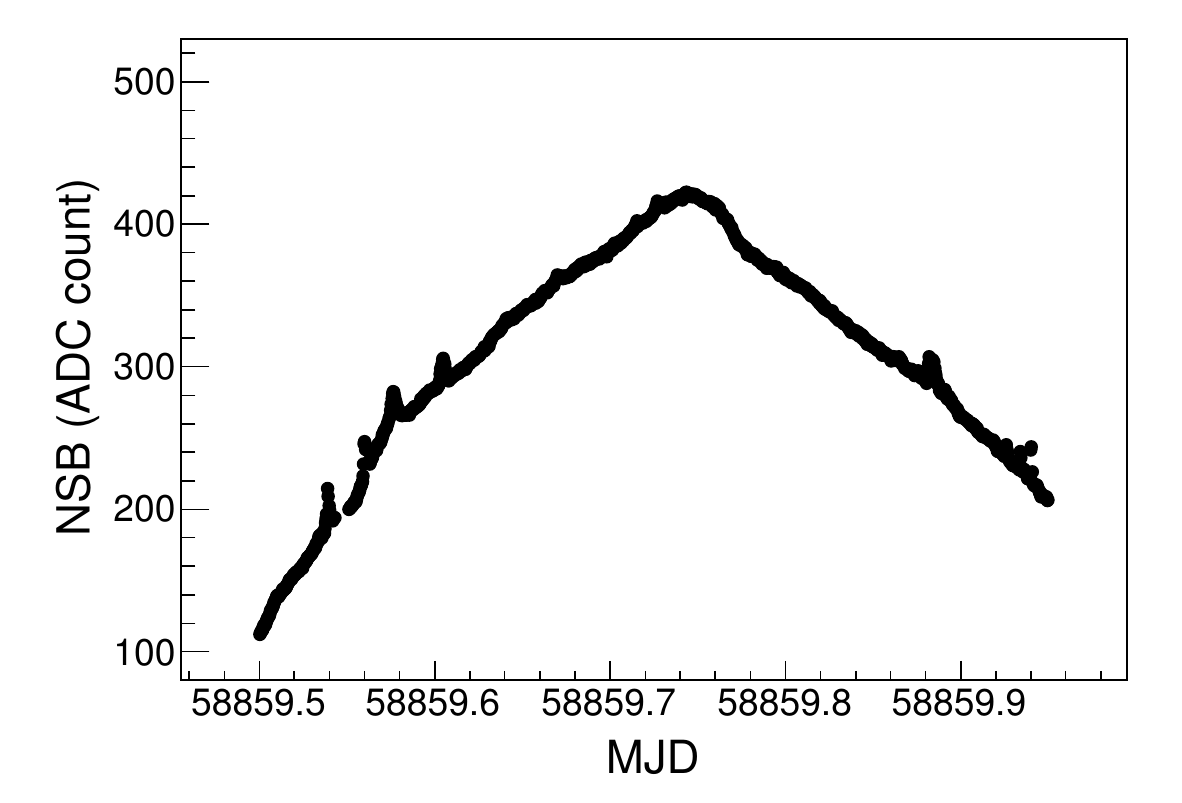} \label{fig:nsb-time-0419}}
	\caption{(a) The elevation angle of the moon as a function of time. (b) The NSB measured by one pixel is modulated by the moonlight. Because of atmospheric scattering and absorption, the amount of moonlight collected by the telescope is a function of the moonlight slant depth in the atmosphere and the scattering angle which is the space angle between the direct moonlight and the telescope. Some small peaks caused by starlight can also be seen on the NSB-time curve.
	}
\end{figure}

The WFCTA telescopes were operated under the full moon on the clear night of January 11, 2020. The elevation angle of the moon as a function of time is shown in Fig.~\ref{fig:moon-trajectory}. The moonlight is scattered by molecules (Rayleigh scattering) and aerosols (Mie scattering), and absorbed in the atmosphere, before it enters the telescope~\cite{Jones_2013,liujiali2018}. Therefore, the measured NSB correlates with the moonlight slant depth in the atmosphere (the elevation angle of the moon). As an example, the NSB measured by one pixel in a camera as a function of time is shown in Fig.~\ref{fig:nsb-time-0419}. The shorter the slant depth (higher elevation angle of the moon), the more moonlight is collected by the telescope. 
The influence of the moon on the SiPM-based Cherenkov telescope has already been measured and studied in FACT~\cite{biland2014}.
The moonlight induces a continuous photo-current in the SiPM, and leads to an additional voltage drop on the bias resistor. Accordingly, the bias voltage, and then the gain, $PDE$, DCR, the crosstalk probability, the afterpulse probability of the SiPM, varies with the intensity of NSB light~\cite{Nagai_2019}.
The continuous photo-current can also generate additional heat on the SiPM.
The above effects can lead to the decrease of the SiPM output charge.
The influence of the moon on the SiPM camera also appears in the WFCTA telescopes. 
The temperature sensor mounted on the SiPM backside can truly reflect the ambient temperature, but it can not truly reflect the heat generated by the SiPM itself, because the heat can be transmitted through the air, the funnel or the connector on the SiPM backside.
Although, it is not easy to study these effects cause by the NSB item by item, the overall effects are discussed in the following. 

\begin{figure}[hbtp]
	\centering
	\includegraphics[width=12cm]{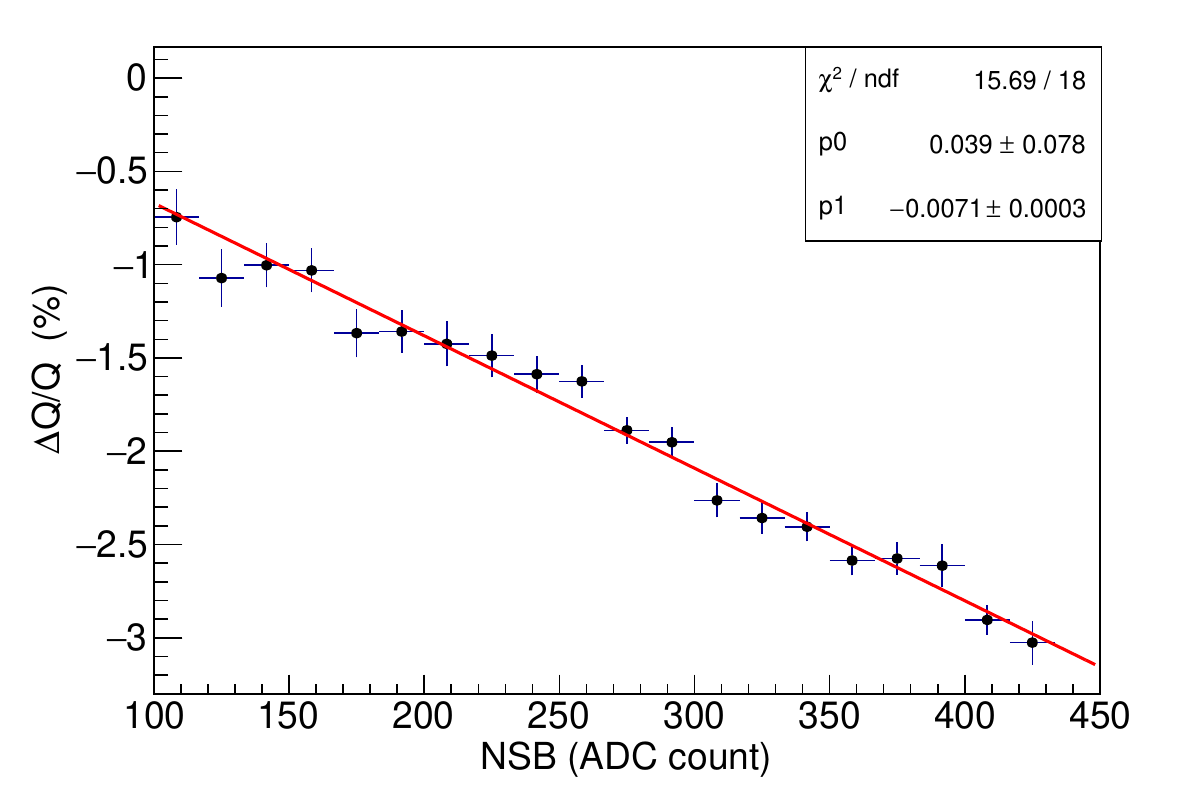}
	\caption{The SiPM relative response charge varies as function of the NSB. As expected, the response charge decreases with increasing NSB because of the SiPM output charge drop introduced by the bias resistor, additional heat and recharging process in the SiPM. The slope is about -0.71\% per 100 ADC counts.}
	\label{fig:deltag_nsb}
\end{figure}

As is shown in Fig.~\ref{fig:deltag_nsb}, the output charge ratio of a SiPM with and without NSB, that is, the relative SiPM response charge, is linearly related to the NSB.
The NSB can be used for an off-line correction of the SiPM response charge reduction due to the moonlight.
After linear fitting of the relationship between the relative response charge and the NSB, the SiPM response charge changes by 0.71\% when the NSB changes by 100 ADC counts, which indicates that the NSB measurement has a sufficient accuracy for the SiPM response charge correction and the correction factor is 0.71\% $\pm$0.03\% per 100 ADC counts. The uncertainty of the NSB measurement comes from the drift of the baseline of the electronic system and the statistical error of the measurement.
By comparing the baseline amplitude with the shutter closed before and after the observation, the drift of the baseline of the electronic system is estimated to be about 13 ADC counts.
More than 80 points of baseline amplitude are recorded in each SiPM signal. The statistical error of the NSB measurement is less than 8 ADC counts in each SiPM signal at the condition of NSB$<$1,000 ADC counts.
So, one of the uncertainties of SiPM response charge corrected by NSB comes from the measurement uncertainty of NSB, which is about 0.15\%. Another SiPM response charge correction uncertainty comes from the measurement uncertainty of the correction factor. The response charge correction uncertainty caused by the correction factor uncertainty is less than 0.3\% at the condition of NSB$<$1,000 ADC counts. The total uncertainty of the SiPM response charge caused by the NSB correction is estimated to be less than 0.45\%.

On average, a certain number of SiPM cells suffer a breakdown and a certain number of SiPM cells are recharged every moment under a strong NSB condition. So, NSB occupies a small part of the dynamic range of the SiPM, e.g. at the maximum NSB shown in Fig.~\ref{fig:deltag_nsb}, the dynamic range of Cherenkov light measurement will be reduced by about 3.4\%.
Due to the increase of the NSB noise on the moon night, the threshold value of each pixel is correspondingly increased to discriminate the signal from the noise.
As a result, the energy threshold of the telescope is increased, e.g. the energy threshold is about 50 TeV on the full moon night. The energy threshold can be further reduced by optimizing the trigger algorithm for the moon night, which is in progress.

\section{Conclusions}\label{sec:sec5}
Eighteen cameras were assembled and tested in a laboratory of Yunnan University, and mounted to telescopes that have been operated for months. As a major component of the LHAASO experiment, the preliminary on-site performance was studied for all the cameras. The most significant improvement of the SiPM-based camera, versus the traditional PMT-based camera is that all of the cameras can be operated even during a full-moon night. The gains of all SiPMs are stabilized within 2\% by bias voltage adjustment and temperature compensation over a typical range of 18~$^{\circ}$C. 
The camera pixels are optimized for maximizing the photon collection area by using the relative smaller SiPMs coupled with square-shaped funnels which fully fill up the FoV of the camera. To achieve the desired performance, tools and jigs were designed and produced to enable successful assembly. Before putting all assembled sub-clusters together as the whole camera, the features of all pixels are calibrated and tested, including the temperature response, the response charge vs. bias voltage, the non-linearity over dynamic range, the signal resolution and DCR. Characterization and construction procedures and corresponding test facilities, such as the 1D-System and the Temp-System, were established. The characterized parameters of all pixels are recorded in a database used in on-line operation and off-line analysis.

Through the past few observational seasons, eighteen telescopes have collected about 100 million cosmic ray events that will be used to measure the energy spectra of cosmic ray protons and other species around 1 PeV. The telescopes can also provide cross calibration of the energy measurement of primary gamma-rays to KM2A in the energy range from 100 TeV to a few PeV. The success summarized here, and the fact that 18,432 SiPMs have been operated regularly, will eventually establish the large-scale application of SiPMs in IACTs.

\section*{Acknowledgment}
The LHAASO project is supported by the National Key R\&D Program of China (2018YFA0404200), the Chinese Academy of Sciences (CAS), the Key Laboratory of Particle Astrophysics, Institute of High Energy Physics, CAS. This work is supported by Yunnan Applied Basic Research Projects (2018FB008, 2018FA004, 2018IC059 and 2018FY001-003), the National Key R\&D Program of China (2018YFA0404204) and the National Natural Science Foundation of China (U1738211, 11675204 and 11905240), Yunnan University, and Grant RTA 6280002 from Thailand Science Research and Innovation. We give thanks to the Swiss Foundation Ernst and Lucie Schmidheiney for having provided the light funnels of the first camera, which were produced according to a technique developed by the University of Geneva. We wish to express our gratitude to the students who helped us fabricating, testing, and operating the SiPM cameras and WFCTA.

\bibliography{mybibfile}

\end{document}